\documentclass[format=acmsmall, review=false]{acmart}

\usepackage{acme20} 
\usepackage{graphicx} 
\usepackage{algorithm}
\usepackage[noend]{algpseudocode}

\title{Frosty: Bringing strong liveness guarantees to the Snow family of consensus protocols.}

\author{Aaron Buchwald}
\affiliation{%
\institution{ \institution{Ava Labs} \country{USA}}
}

\author{Stephen Buttolph}
\affiliation{%
\institution{ \institution{Ava Labs}  \country{USA}}
}

\author{Andrew Lewis-Pye}
\affiliation{%
\institution{ \institution{London School of Economics}  \country{UK}}
}

\author{Patrick O'Grady}
\affiliation{%
\institution{ \institution{Ava Labs}  \country{USA}}
}

\author{Kevin Sekniqi}
\affiliation{%
\institution{ \institution{Ava Labs}  \country{USA}}
}

\date{March 2024}

\begin{document}

\begin{abstract}
    Snowman is the consensus protocol implemented by the Avalanche blockchain and is part of the Snow family of protocols, first introduced in the Avalanche whitepaper \cite{rocket2019scalable}. A major advantage of Snowman is that each consensus decision only requires an expected constant communication overhead per processor in the `common' case that the protocol is not under substantial Byzantine attack, i.e.\ it provides a solution to the scalability problem which ensures that the expected communication overhead per processor is independent of the total number of processors $n$ during normal operation. This is the key property that would enable a consensus protocol to scale to 10,000 or more independent validators (i.e. processors). On the other hand, the two following concerns have remained: 
    \begin{enumerate}
        \item Providing formal proofs of consistency for Snowman has presented a formidable challenge. 
        \item Liveness attacks exist in the case that a Byzantine adversary controls more than $O(\sqrt{n})$ processors, slowing termination to more than a logarithmic number of steps.
    \end{enumerate}
    In this paper, we address the two issues above. We consider a Byzantine adversary that controls at most $f<n/5$ processors. First, we provide a simple proof of consistency for Snowman. Then we supplement Snowman with a `liveness module' that can be triggered in the case that a substantial adversary launches a liveness attack, and which guarantees liveness in this event by \emph{temporarily} forgoing the communication complexity advantages of Snowman, but without sacrificing these low communication complexity advantages during normal operation.   
\end{abstract}

\maketitle

\section{Introduction} \label{intro}
Recent years have seen substantial interest in developing consensus protocols that work efficiently at scale. In concrete terms, this means looking to minimize the latency and communication complexity per consensus decision as a function of the number of processors (participants/validators) $n$. The  Dolev-Reischuk bound~\cite{dolev1985bounds}, which asserts that deterministic protocols require $\Omega(n^2)$ communication complexity per consensus decision, presents a fundamental barrier in this regard: deterministic protocols that can tolerate a Byzantine (i.e. arbitrary) adversary of size $O(n)$ must necessarily suffer a quadratic blow-up in communication cost as the size of the network grows. It is precisely this relationship that makes these protocols susceptible to considerable slowdown when a high number of processors is present.

\vspace{0.2cm}
\noindent \textbf{Probabilistic sortition}. One approach to dealing with this quadratic blow-up in communication cost, as employed by protocols such as Algorand \cite{chen2016algorand}, is to utilize probabilistic \emph{sortition} \cite{king2011breaking,abraham2019communication}. Rather than have \emph{all} processors participate in every consensus decision, the basic idea is to sample a \emph{committee} of sufficient size that the proportion of Byzantine committee members is almost certainly close to the proportion of all processors that are Byzantine.  Sampled committees of constant bounded size can then be used to implement consensus, thereby limiting the communication cost. In practical terms, however, avoiding Byzantine control of committees requires each committee to have a number of members sufficient that the \emph{quadratic communication cost for the committee} is already substantial, e.g.\ Algorand requires committees with $k$ members, where $k$ is of the order of one thousand, meaning that $k^2$ is already large.

\vspace{0.2cm}
\noindent \textbf{The Snow family of consensus protocols}. In \cite{rocket2019scalable}, a family of consensus protocols was specified, providing an alternative approach to limiting communication costs. These protocols are all based on a common approach that is best described by considering a binary decision game. For the sake of simplicity, let us initially consider the Snowflake protocol\footnote{In \cite{rocket2019scalable}, other variants such as the Slush and Snowball protocols are also described.}, which uses three parameters: $k$, $\alpha>k/2$, and $\beta$ (for the sake of concreteness, in this paper we will focus on the example that $k=80$). Suppose that each processor begins with an initial color, either red or blue. Each processor $p$ then proceeds in rounds. In each round, $p$ randomly samples $k$ processors from the total population and asks those processors to report their present color. If at least $\alpha$ of the reported values are the opposite of $p$'s present color, then $p$ adopts that opposite color.  If $p$ sees $\beta$ consecutive rounds in which at least $\alpha$ of the reported values are red,  then $p$ decides red (and similarly for blue). 

\vspace{0.2cm} The outcome of this dynamic sampling process can be informally described as follows when the adversary is sufficiently bounded (a formal analysis for a variant of Snowflake that we call Snowflake$^+$ is given in Section \ref{Snowflakeanalysis}). Once the proportion of the population who are red, say, passes a certain tipping point, it holds with high probability that the remainder of the (non-Byzantine) population will quickly become red (and symmetrically so for blue). If $\beta$ is set appropriately, then the chance that any correct processor decides on red before this tipping point is reached can be made negligible, meaning that once any correct processor decides on red (or blue), they can be sure that all other correct processors will quickly decide the same way. The chance that correct processors decide differently can thus be made negligible through an appropriate choice of parameter values. If correct processors begin heavily weighted in favor of one color, then convergence on a decision value  will happen very quickly, while variance in random sampling is required to tip the population in one direction in the case that initial inputs are evenly distributed. 

\vspace{0.2cm} While the discussion above considers a single binary decision game, the `Snowman' protocol, formally described and analysed for the first time in this paper, shows that similar techniques can be used to efficiently solve State Machine Replication (SMR) \cite{schneider1990implementing}. The transition from simple consensus (Byzantine Agreement \cite{lamport1982byzantine}) to an efficient SMR protocol is non-trivial, and is described in detail in Sections \ref{Snowman} and \ref{Snowmansecurity}.  A major benefit of the approach is that it avoids the need for all-to-all communication. In an analysis establishing that there is only a small chance of consistency failure, the value of $k$ can be specified independent of $n$, and each round requires each processor to collect reported values from only $k$ others.

\vspace{0.2cm} 
\noindent \textbf{Our contribution}.  The Snowman protocol is presently used by the Avalanche blockchain to implement SMR. However, the two following concerns have remained: 
    \begin{enumerate}
        \item Providing formal proofs of consistency for Snowman has presented a formidable challenge. 
        \item Liveness attacks exist in the case that a Byzantine adversary controls more than $O(\sqrt{n})$ processors \cite{rocket2019scalable}, meaning that finalization is no longer guaranteed to occur in a logarithmic number of steps.  
    \end{enumerate}
    In this paper, we consider a Byzantine adversary that controls at most $f<n/5$ processors, and address the two issues above. With respect to issue (1): 
    \begin{itemize} 
    \item We describe a variant of Snowflake, called Snowflake$^+$. 
    \item For appropriate choices of parameter values, we  give a simple proof that Snowflake$^+$ satisfies `validity' and `agreement' except with small error probability. 
    \item We give a complete specification of a version of Snowman that builds on Snowflake$^+$. This is the first formal description of the Snowman protocol. 
    \item For appropriate choices of parameter values, we give a simple proof that the resulting Snowman protocol satisfies consistency except with small error probability.  
        \item We also describe a variant of Snowflake$^+$ called Error-driven Snowflake$^+$, that can be used to give very low latency in the `common case'.
    \end{itemize}

    \vspace{0.2cm} With regard to issue (2),  we note that malicious liveness attacks on Avalanche have not been observed to date. It is certainly desirable, however, to have strong guarantees in the case that a large adversary launches an attack on liveness. 
    The approach we take in this paper is therefore to strike a practical balance. More specifically, we aim to specify a protocol that is optimised to work efficiently in the `common case' that there is no substantial Byzantine attacker, but which also provides a `fallback' mechanism in the worst case of a substantial attack on liveness.
    To this end, we then describe how to supplement Snowman with a `liveness module'. The basic idea is that one can use Snowman to reach fast consensus under normal operation, and can then trigger an `epoch change' that temporarily implements some standard quorum-based protocol to achieve liveness in the case that a substantial adversary attacks liveness. In the (presumably rare) event that a substantial adversary attacks liveness, liveness is thus ensured by \emph{temporarily} forgoing the communication complexity advantages of Snowman during normal operation. The difficulty in implementing such a module is to ensure that interactions between the different modes of operation do not impact consistency. We give a formal proof that the resulting protocol, called Frosty, is consistent and live, except with small error probability. 

    \vspace{0.2cm}
\noindent \textbf{Paper structure}. Section \ref{model} describes the formal model. Section \ref{simp} describes Snowflake$^+$ and gives pseudocode for the protocol. Section \ref{Snowflakeanalysis} gives a  proof of agreement and validity for Snowflake$^+$ and describes Error-driven Snowflake$^+$. 
Section \ref{Snowman} describes the Snowman protocol, including pseudocode. Section \ref{Snowmansecurity} gives a proof of consistency for Snowman. Section \ref{Avalanche2} describes the liveness module and gives pseudocode for the resulting protocol, called Frosty. Section \ref{Avalanche2analysis} proves liveness and consistency for Frosty.

\section{The model} \label{model}

We consider a set $\Pi= \{ p_0,\dots, p_{n-1} \}$ of $n$ processors. Processor $p_i$ is told $i$ as part of its input. For the sake of simplicity, we assume a static adversary that controls up to $f$ of the processors, where $f$ is a known bound. Generally, we will assume $f<n/5$. The bound $f<n/5$ is chosen only so as to give as simple a proof as possible in Section \ref{Snowflakeanalysis}, and providing an analysis for larger $f$ is the subject of future work. 
 A processor that is controlled by the adversary is referred to as \emph{Byzantine}, while processors that are not Byzantine are \emph{correct}. Byzantine processors may display arbitrary behaviour, modulo our cryptographic assumptions (described below).

\vspace{0.2cm} 
\noindent \textbf{Cryptographic assumptions}. Our cryptographic assumptions are standard for papers in distributed computing. Processors communicate by point-to-point authenticated channels. We use a cryptographic signature scheme, a public key infrastructure (PKI) to validate signatures, and a collision-resistant hash function $H$. 
 We assume a computationally bounded adversary. Following a common standard in distributed computing, and for simplicity of presentation (to avoid the analysis of certain negligible error probabilities), we assume these cryptographic schemes are perfect, i.e.\ we restrict attention to executions in which the adversary is unable to break these cryptographic schemes. In a given execution of the protocol, hash values are thus assumed to be unique.

 \vspace{0.2cm} 
\noindent \textbf{Communication}. As noted above, processors communicate using point-to-point authenticated channels. We consider the standard synchronous model: for some known bound $\Delta$, a message sent at time $t$ must arrive by time $t + \Delta$.  

 \vspace{0.2cm} 
\noindent \textbf{The binomial distribution}. Consider $k$ independent and identically distributed random variables, each of which has probability $x$ of taking the value `red'. We let $\text{Bin}(k,x,m)$ denote the probability that $m$ of the $k$ values are red, we write $\text{Bin}(k,x,\geq m)$ to denote the probability that \emph{at least} $m$ values are red (and similarly for $\text{Bin}(k,x,\leq m)$).

 \vspace{0.2cm} 
\noindent \textbf{Dealing with small probabilities}. In analysing the security of a cryptographic protocol, one standardly regards a function $f:\mathbb{N} \rightarrow \mathbb{N}$ as \emph{negligible} if, for every $c\in \mathbb{N}$, there exists $N_c\in \mathbb{N}$ such that, for all $x\geq N_c$, $|f(x)|<1/x^c$. Our concerns here, however, are somewhat different. As noted above, we assume the cryptographic schemes utilized by our protocols are perfect. For certain \emph{fixed} parameter values (e.g.\ setting $n=500$, $k=80$, $\alpha=41$ and $\beta=12$ in an instance of Snowflake, as described in Section \ref{intro}), we want to be able to argue that error probabilities are sufficiently small that they can reasonably be dismissed.

In our analysis, we will therefore identify certain events as occurring with \emph{small} probability (e.g. with probability $<10^{-20}$), and may then condition on those events not occurring. 
Often, we will consider specific events, such as the probability in a round-based protocol that a given processor performs a certain action $x$ in a given round. In dismissing small error probabilities, one then has to take account of the fact that there may be many opportunities for an event of a given type to occur, e.g. any given processor may perform action $x$ in any given round. How reasonable it is to condition on no correct processor performing action $x$ may therefore depend on the number of processors and the number of rounds, and we assume `reasonable' bounds on these values. As an example, consider the Snowflake protocol, as described in Section \ref{intro}, and suppose $k=80$ and that at most 1/5 of the processors are Byzantine. Suppose that, at the beginning of a certain round, at least 75\% of the correct processors are red. Then a calculation for the binomial distribution shows that the probability a correct processor receives at least 72 blue responses from the 80 processors it samples in that round is upper bounded by $1.18\times 10^{-20}$,i.e.\ $\text{Bin}(80,0.2+(0.8 \times 0.25),\geq 72)<1.18 \times 10^{-20}$.  To upper bound the probability that there exists \emph{any} round in which at least 75\% of correct processors are red and \emph{some} correct processor receives at least 72 blue responses, we just apply the union bound. For the sake of concreteness, suppose that at most 10,000 processors run the protocol for at most 1000 years, executing at most 5 rounds a second. This means that less than $1.6\times 10^{11}$ rounds are executed. Since there are at most 10,000 processors, the union bound thus gives a cumulative error probability less than $(1.18\times 10^{-20}) \times 10^4 \times (1.6\times 10^{11})<2\times 10^{-5}$. We will address such accountancy issues as they arise.

We stress that accounting for \emph{small} error probabilities in the manner described above (rather than showing error probabilities are negligible functions of the parameter inputs) also allows us to give particularly straightforward security proofs for Snowflake$^+$, Snowman, and Frosty.

 \vspace{0.2cm} 
\noindent \textbf{A comment on the use of synchrony}. We simplify our analysis by having correct processors execute the protocol executions in cleanly defined \emph{rounds}. Each correct processor thus samples the values of some others in round 1, before adjusting local values based on that sample. All correct processors then proceed to round 2, and so on. As discussed in In Section \ref{fincom},  a follow-up paper will show how the analysis here can be extended to deal with a \emph{responsive} version of the protocol in which each correct processor can proceed through rounds as fast as local message delays allow, i.e. a processor may proceed to round $s+1$ as soon as they receive sufficiently many responses for round $s$.

\section{A simple protocol for Byzantine Agreement:  Snowflake$^{+}$.} \label{simp}

We begin by describing a simple probabilistic protocol for binary Byzantine Agreement, called Snowflake$^+$, which will act as a basic building block for the Snowman protocol (described later in Section \ref{Snowman}). While a similar analysis could be given for Snowflake (as described in the original whitepaper \cite{rocket2019scalable}), an advantage of Snowflake$^+$ is that it allows for a simpler proof to establish that error probabilities are small and, as described in further detail in Section \ref{flex},  Snowflake$^+$ is also  easily adapted to give flexible termination conditions, giving low latency in the good case that most processors are correct. Similar considerations also apply when comparing Snowflake$^+$ and Snowball (also described in the original whitepaper \cite{rocket2019scalable}).


\vspace{0.2cm}
\noindent \textbf{The inputs}. Each processor $p_i$ begins with a value $\mathtt{input}_i\in \{ 0, 1 \}$.  

\vspace{0.2cm}
\noindent \textbf{The requirements}. A probabilistic protocol for Byzantine agreement is required to satisfy the following properties, except with small error probability: 

\noindent \emph{Agreement}: No two correct processors output different values.   

\noindent \emph{Validity}: If every correct processor $i$ has the same value $\mathtt{input}_i$, then no correct processor outputs a value different than this common input.

\noindent \emph{Termination}: Every correct processor gives an output.  

\vspace{0.2cm}
\noindent \textbf{Recalling Snowflake}. Since Snowflake$^+$ is a simple variant of Snowflake, let us first informally recall the Snowflake protocol. Snowflake uses three parameters: $k$, $\alpha>k/2$, and $\beta$. Each processor $p_i$ maintains a variable $\mathtt{val}_i$, initially set to $\mathtt{input}_i$. The instructions proceed in rounds.  In each round, processor $p_i$ randomly samples $k$ processors from the total population and asks each of those processors $p_j$ to report their present value $\mathtt{val}_j$. If at least $\alpha$ of the reported values are the opposite of $p_i$'s present value $\mathtt{val}_i$, then $p_i$ sets $\mathtt{val}_i:=1-\mathtt{val}_i$.  If $p_i$ sees $\beta$ consecutive rounds in which at least $\alpha$ of the reported values are 1,  then $p_i$ decides 1 (and similarly for 0).

\vspace{0.2cm} Snowflake$^+$ is similar to Snowflake, except that we now use two parameters $\alpha_1$ and $\alpha_2$, rather than a single parameter $\alpha$. 

\vspace{0.2cm}
\noindent \textbf{The protocol parameters for Snowflake$^+$}. The protocol parameters are $k,\alpha_1,\alpha_2,\beta \in \mathbb{N}_{>0}$ and satisfy the constraints that $\alpha_1>k/2$ and $\alpha_2\geq \alpha_1$. Each processor $p_i$ also maintains a variable $\mathtt{val}_i$, initially set to $\mathtt{input}_i$. The parameter $k$ determines sample sizes. The parameter $\alpha_1$ is used to determine when processor $p_i$ changes $\mathtt{val}_i$. Parameters $\alpha_2$ and $\beta$ are used to determine the conditions under which $p_i$ will output and terminate.   

\vspace{0.2cm}
\noindent \textbf{The protocol instructions for Snowflake$^+$}. The instructions are divided into rounds, with round $s$ occurring at time $2\Delta s$. In round $s$, processor $p_i$:
\begin{enumerate} 
\item Sets $\langle p_{1,s},\dots p_{k,s} \rangle$ to be a sequence of $k$ processors (specific to $p_i$). For $j\in [1,k]$,  $p_{j,s}$ is sampled from the uniform distribution\footnote{In proof-of-stake implementations, sampling will be stake-weighted, but, for the sake of simplicity of presentation, we ignore such issues here.} on all processors (so sampling is ``with replacement''). 
\item Requests each $p_{j,s}$ (for $j\in [1,k]$) to report its present value $\mathtt{val}_j$.  
\item Waits time $\Delta$ and reports its present value $\mathtt{val}_i$ to any processor that has requested it in round $s$.
\item Waits another $\Delta$ and considers the values reported in round $s$:
\begin{itemize}
 \item If at least $\alpha_1$ of the reported values are $1-\mathtt{val}_i$, then $p_i$ sets $\mathtt{val}_i:= 1-\mathtt{val}_i$.
 \item If $p_i$ has seen $\beta$ consecutive rounds in which at least $\alpha_2$ of the reported values are equal to $\mathtt{val}_i$, then $p_i$ outputs this value and terminates.
\end{itemize}

\end{enumerate}

The pseudocode is described in Algorithm 1.

 \vspace{0.2cm}
 \noindent In Section \ref{Snowflakeanalysis}, we will show that Snowflake$^+$ satisfies agreement and validity for appropriate choices of the protocol parameters, and so long as $f<n/5$.  
 We do not give a formal analysis of termination for Snowflake$^+$: Once Snowflake$^+$ has been used to define Snowman in Section \ref{Snowman}, in Section \ref{Avalanche2} we will describe  how to augment Snowman with a liveness module (guaranteeing termination), which is formally analysed in Section \ref{Avalanche2analysis}.

\begin{algorithm} \label{pc:Snowflake}
\caption{Snowflake$^{+}$: The instructions for processor $p_i$}
\begin{algorithmic}[1]

    \State \textbf{Inputs} 
    \State $\mathtt{input}_i\in \{ 0, 1 \}$              \Comment $p_i$'s input
    \State $\Delta, k,\alpha_1,\alpha_2,\beta \in \mathbb{N}$  \Comment Protocol parameters

    \State \textbf{Local variables} 
    
   \State $\mathtt{val}_i$, initially set to $\mathtt{input}_i$   \Comment $p_i$'s present `value'      
    \State $\mathtt{count}$, initially set to $0$ \Comment Output once $\mathtt{count}$ reaches $\beta$
    \State $v_i(j,s)$, initially undefined           \Comment Stores at most one received value per round
     
    \State 

  \State \textbf{The instructions for round $s$, beginning at time $2\Delta s$:}

  \State \hspace{0.3cm} Form sample sequence $\langle p_{1,s},\dots p_{k,s} \rangle$;           \Comment Sample  with replacement

  \State \hspace{0.3cm} For $j\in [1,k]$, send $s$ to $p_{j,s}$;    \Comment Ask  $p_{j,s}$ for present value

   \State \hspace{0.3cm}  Wait $\Delta$;

   \State \hspace{0.3cm} For each $j$ such that $p_i$ has received $s$ from $p_j$:

   \State \hspace{0.8cm} Send $(s,\mathtt{val}_i)$ to $p_j$;

   \State \hspace{0.3cm}  Wait $\Delta$;

   \State \hspace{0.3cm} For each $j\in [1,k]$:
   \State \hspace{0.8cm} \textbf{If} $p_i$ has received a first message $(s,v)$ from $p_{j,s}$;
   \State \hspace{1.3cm}  Set $v_i(j,s):=v$;
   \State \hspace{0.8cm} \textbf{Else} set $v_i(j,s):= \bot$;

   \State \hspace{0.3cm} \textbf{If} $|\{ j: 1\leq j \leq k, v_i(j,s)==1-\mathtt{val}_i \}|\geq \alpha_1$, set $\mathtt{val}_i:=1-\mathtt{val}_i$, $\mathtt{count}:=0$;

  \State \hspace{0.3cm} \textbf{If} $|\{ j: 1\leq j \leq k, v_i(j,s)==\mathtt{val}_i \}|< \alpha_2$, set $\mathtt{count}:=0$;

  \State \hspace{0.3cm} \textbf{If} $|\{ j: 1\leq j \leq k, v_i(j,s)==\mathtt{val}_i \}| \geq  \alpha_2$, set $\mathtt{count}:=\mathtt{count}+1$;

  \State \hspace{0.3cm} \textbf{If} $\mathtt{count}\geq \beta$, output $\mathtt{val}_i$ and terminate.

\end{algorithmic}
\end{algorithm}



\section{Security analysis of Snowflake$^+$} \label{Snowflakeanalysis}

We assume $f<n/5$. For the sake of concreteness, we establish satisfaction of agreement and validity for $k=80$, $\alpha_1=41$, $\alpha_2=72$, and $\beta=12$, under the assumption that the population size $n\geq 500$. We make the assumption that $f<n/5$ and $n\geq 500$ only so as to be able to give as simple a proof as possible: a more fine-grained analysis for smaller $n$ is the subject of future work. 
 

\vspace{0.2cm}
\noindent \textbf{Coloring the processors}. Since 0 and 1 are not generally used as adjectives, let us say a correct processor $p_i$ is `blue' in round $s$ if $\mathtt{val}_i=0$ at the beginning of round $s$, and that $p_i$ is `red' in round $s$ if $\mathtt{val}_i=1$ at the beginning of round $s$. Recall (from Algorithm 1) that $v_i(j,s)$ is the color that $p_j$ reports to $p_i$ in round $s$.  We'll say a correct processor $p_i$ `samples $x$ blue' in round $s$ if $|\{ j: 1\leq j \leq k, v_i(j,s)=0 \}|=x$ (and similarly for red).  We'll also extend this terminology in the obvious way, by saying that a processor outputs `blue' if it outputs 0 and outputs `red' if it outputs 1. In the below, we'll focus on the case that, in the first round in which a correct processor outputs (should such a round exist), some correct processor outputs red. A symmetric argument can be made for blue.

\vspace{0.2cm} In the following argument, we will adopt the conventions described in Section \ref{model} concerning the treatment of small error probabilities. We will identify certain events as occurring with \emph{small} probability (e.g. with probability $<10^{-20}$), and may then condition on those events not occurring. Where there are multiple opportunities for an event of a certain type to occur, one must be careful to account for the accumulation of small error probabilities. To deal with the accumulation of small error probabilities, we suppose that at most 10,000 processors execute the protocol for at most 1000 years, executing at most 5 rounds per second. 

\vspace{0.2cm}
\noindent \textbf{Establishing Agreement}. The argument consists of four parts: 

\vspace{0.2cm} 
\noindent \textbf{Part 1}. First, let us consider what happens when the proportion of correct processors that are red reaches a certain threshold. In particular, let us consider what happens when at least 75\% of the correct  processors are red in a given round $s$. A direct calculation for the binomial distribution shows that the probability a given correct processor is red in round $s+1$ is then at least 0.9555, i.e.\ $\text{Bin}(80,0.8 \times 0.75,\geq41)>0.9555$. Assuming a population of at least 500, of which at least 80\%  are correct (meaning that at least 400 are correct), another direct calculation for the binomial distribution shows that the probability that it fails to be the case that more than 5/6 of the correct processors are red in round $s+1$ is upper bounded by $1.59\times 10^{-20}$, i.e.\ $\text{Bin}(n,0.9555,\leq 5n/6)<1.59 \times 10^{-20}$ for $n\geq 400$. 
Note that this argument requires no knowledge as to the state of each processor in round $s$, other than the fact that at least 75\% of the correct  processors are red. 

The analysis above applies to any single given round $s$. Next, we wish to iterate the argument over rounds in order to bound the probability that the following statement holds for \emph{all} rounds: 
\begin{enumerate} 
\item[$(\dagger_1)$] If at least 75\% of the correct processors are red in any round $s$, then, in all rounds $s'$ with $s'>s$, more than 5/6 of the correct processors are red.
\end{enumerate}
\noindent To bound the probability that $(\dagger_1)$ fails to hold, we can bound the number of rounds, and then apply the union bound to our analysis of the error probability in each round. Suppose that the protocol is executed for at most 1000 years, with at most 5 rounds executed per second.  This means that less than $1.6\times 10^{11}$ rounds are executed. The union bound thus gives a cumulative error probability of less than $(1.6\times 10^{11}) \times (1.59\times 10^{-20})  < 3\times 10^{-9}$, meaning that $(\dagger_1)$ fails to hold  with probability at most  $3\times 10^{-9}$. 

\vspace{0.2cm} 
\noindent \textbf{Part 2}. 
A  calculation for the binomial distribution shows that if at least 75\% of correct processors are red in a given round $s$, then the probability that a given correct processor $p_i$ samples at least 72 blue in round $s$ is upper bounded by $1.18\times 10^{-20}$, i.e.\ $\text{Bin}(80,0.2+(0.8 \times 0.25),\geq 72)<1.18 \times 10^{-20}$. If at most 10,000 processors execute the protocol for at most 1000 years, executing at most 5 rounds per second, we can then apply the union bound to conclude that the following statement fails to hold with probability at most $(1.18\times 10^{-20})\times 10000 \times (1.6 \times 10^{11})<  2\times 10^{-5}$:

\begin{enumerate}
\item[$(\dagger_2)$] If at least 75\% of the correct processors are red in any round $s$, then no correct processor samples at least 72 blue in round $s$.
\end{enumerate}

\vspace{0.2cm} 
\noindent \textbf{Part 3}.  Another direct calculation for the binomial distribution shows that, if \emph{at most} 75\% of correct processors are red in a given round $s$, then the probability a given correct processor $p$ samples 72 or more red in round $s$ is upper bounded by 0.0131, i.e.\ $\text{Bin}(80,(0.75\times 0.8)+0.2,\geq 72)<0.0131$. If, for some $x\geq 1$ it then holds that at most 75\% of correct processors are red in round $s+x$, then (independent of previous events), the probability $p$ samples 72 or more red in round $s+x$ is again upper bounded by 0.0131.  So, if we consider any 12 given consecutive rounds and any given correct processor $p$, the probability that at most 75\% of correct processors are red and $p$ samples at least 72 red in all 12 rounds is upper bounded by  $0.0131^{12}<10^{-22}$. If at most 10,000 processors execute the protocol for at most 1000 years, executing at most 5 rounds per second, we can then apply the union bound to conclude that the following statement fails to hold with probability at most $10^{-22}\times 10000 \times (1.6 \times 10^{11})<  2\times 10^{-7}$:

\begin{enumerate}
\item[$(\dagger_3)$] If a correct processor outputs red in some round $s+11$, then, for at least one round $s'\in [s,s+11]$, at least 75\% of correct processors are red in round $s'$.
\end{enumerate}

\vspace{0.2cm} 
\noindent \textbf{Part 4}. Now we put parts 1--3 together. From the union bound and the analysis above, we may conclude that $(\dagger_1)-(\dagger_3)$ all hold, except with probability at most 
$(3\times 10^{-9})+(2\times 10^{-5})+( 2\times 10^{-7})<3\times 10^{-5}$. So, suppose that $(\dagger_1)-(\dagger_3)$ all hold. 
According to $(\dagger_3)$,  if a correct processor is the (potentially joint) first to output and outputs  red after sampling in round $s+11$, at least one round $s'\in [s,s+11]$ must satisfy the condition that at least 75\% of correct processors are red in round $s'$. From $(\dagger_1)$, it follows that at least 5/6 of the correct processors must be red in all rounds $>s'$.    From $(\dagger_2)$, it follows that no correct processor ever outputs blue. This suffices to show that Agreement is satisfied, except with small  error probability. 

\vspace{0.2cm}
\noindent \textbf{Establishing Validity}. A similar (but even simpler) argument suffices to establish validity. Suppose that all honest nodes have the same input, red say (i.e. 1). By the same reasoning as above, since round 0 satisfies the condition that at least 75\% (in fact 100\%) of correct processors are red, the following statement fails to hold  with probability at most $3\times 10^{-9}$: 

\begin{enumerate} 
\item[$(\dagger_4)$] In every round, more than 5/6 of the correct processors are red.
\end{enumerate}

\noindent From $(\dagger_2)$ and $(\dagger_4)$ it follows that no correct processor outputs blue, as required.

\vspace{0.2cm}
\noindent \textbf{Dealing with different parameter values}. The argument above is easily adapted to deal with alternative parameter values. If we fix $\alpha_1:=\lfloor k/2 \rfloor +1$, then error probabilities will be smaller for larger values of $\alpha_2$ and $\beta$. For smaller values of $\alpha_2$, similar error probabilities can be obtained by increasing $\beta$ -- the required values for $\beta$ are easily found by adapting the binomial calculations above. Examples are given in Section \ref{flex}.

\subsection{Error-driven  Snowflake$^+$} \label{flex} 
In Section \ref{Snowflakeanalysis}, we considered a fixed value $\alpha_2=72$, for $k=80$. While considering a fixed $\alpha_2$ suffices for the analysis there, it is also useful to simultaneously consider multiple values of $\alpha_2$, giving rise to a number of different conditions for termination.  Considering a range of simultaneous termination conditions for different values of $\alpha_2$ serves two functions: Considering lower values of $\alpha_2$ allows one to deal with a greater percentage of offline/faulty processors, while higher values of $\alpha_2$ give quick decision conditions and low latency in the good case.

\vspace{0.2cm} 
 \emph{Error-driven Snowflake$^+$} is the same as Snowflake$^+$, except that one simultaneously considers multiple possible values of $\alpha_2\leq k$. Each $\alpha_2$ now gives rise to a different $\beta$ that determines the conditions for termination. 
 The corresponding values are shown in Table \ref{table1}.

 \vspace{0.2cm}
\noindent \textbf{How the values in Table \ref{table1} are calculated}. In Section \ref{Snowflakeanalysis}, it was $(\dagger_3)$ which played a crucial role in establishing the relationship between $\alpha_2$ and $\beta$ for a given error probability $\epsilon >0$. Assuming that at most 75\% of correct processors are red, a calculation for the binomial distribution then upper bounds the probability $p$ that a given correct processor samples at least $\alpha_2$ red in a given round. For a given error probability $\epsilon$, the corresponding $\beta$ shown in Table \ref{table1} is the least integer such that $p^\beta<\epsilon$. The value $p^\beta$ upper bounds the probability of a given correct processor sampling at least $\alpha_2$ red in $\beta$ given consecutive rounds, under the assumption that at most 75\% of correct processors are red in each round. 

\vspace{0.2cm}
Table \ref{table1} also shows how $\beta$ depends on $\alpha_2$ for larger error bounds ($\epsilon<10^{-14}$ and $\epsilon<10^{-6}$). Correct processors may use the corresponding lower values of $\beta$ in the case that they are willing to accept higher error probabilities for the sake of achieving low latency, i.e.\ terminating in a small number of rounds. 

\vspace{0.2cm} 
\noindent \textbf{Low latency in standard operation}. Analysis of data from the Avalanche blockchain shows that, at any given point in time, one can expect close to 100\% of contributing processors to act correctly. For Error-driven Snowflake$^+$, this corresponds to a scenario where the vast majority of processors are correct, and where initial inputs are generally highly biased in favor of one color.  The conditions in Table \ref{table1} that allow for quick termination (using $\beta=3$, 4 or 5, say) can therefore be expected to be commonly satisfied, and give a significant improvement in latency  for the standard case that most processors act correctly.  

\vspace{0.2cm} 
\noindent \textbf{The accumulation of error probabilities}. Accepting multiple conditions for termination gives an overall error probability that can be (generously) upper bounded simply by applying the union bound. In Table 1, 16 different termination conditions are listed. If processors apply all of these termination conditions simultaneously, then this will lead to at most a 16-fold increase in error probability.

\begin{center}
\begin{table}
\begin{tabular}{ |c|c|c|c| } 
 \hline
 $\alpha_2$ & $\beta$ for & $\beta$ for & $\beta$ for\\ 
            & $\epsilon<10^{-22}$ & $\epsilon<10^{-14}$ & $\epsilon<10^{-6}$ \\
\hline             
 80 & 3 & 2 & 1 \\ 
79 & 4 & 3 & 1 \\ 
78 & 5 & 3 & 2 \\ 
77 & 5 & 4 & 2 \\ 
76 & 6 & 4 & 2 \\ 
75 & 7 & 5 & 2 \\ 
74 & 9 & 6 & 3 \\ 
73 & 10 & 7 & 3 \\ 
72 & 12 & 8 & 4 \\ 
71 & 15 & 10 & 4 \\ 
70 & 18 & 12 & 5 \\ 
69 & 23 & 15 & 7 \\ 
68 & 29 & 18 & 8 \\
67 & 37 & 24 & 10 \\ 
66 & 48 & 31 & 14 \\ 
65 & 65 & 41 & 18 \\ 
 \hline
\end{tabular}

\vspace{0.3cm} 
\caption{The required $\beta$ as a function of $\alpha_2$ and the error bound.}
\label{table1} 
\end{table}
\end{center}

\section{The Snowman protocol} \label{Snowman}

Since the Snowman protocol is not specified in the original whitepaper \cite{rocket2019scalable}, we give a precise description and pseudocode here. 

\subsection{Transactions and blocks}
To specify a protocol for State-Machine-Replication (SMR), we suppose processors are sent (signed) transactions during the protocol execution: Formally this can be modeled by having processors be sent transactions by an \emph{environment}, e.g.\ as in \cite{lewis2023permissionless}. Processors may use received transactions to form \emph{blocks} of transactions. To make the analysis as general as possible, we decouple the process of block production from the core consensus engine. We therefore suppose that some given process for block generation operates in the background, and that valid blocks are gossiped throughout the network. We do not put constraints on the block generation process, and allow that it may produce equivocating blocks, etc. In practice, block generation could be specified simply by having a rotating sequence of leaders propose blocks, or through a protocol such as Snowman$^{++}$, as actually used by the present implementation of the Avalanche blockchain (for a description of Snowman$^{++}$, see \cite{plusplus}). 

\vspace{0.2cm} 
\noindent \textbf{Blockchain structure}. We consider a fixed \emph{genesis block} $b_0$. In a departure from the approach described in the original Avalanche whitepaper \cite{rocket2019scalable}, which built a directed acyclic graph (DAG) of blocks, we consider a standard blockchain architecture in which each block $b$ other than $b_0$ specifies a unique \emph{parent}.  
If $b'$ is the parent of $b$, then $b$ is referred to as a \emph{child} of $b'$. In this case, the ancestors of $b$ are $b$ and any ancestors of $b'$. Every block must have $b_0$ as an ancestor. The descendants of any block $b$ are $b$ and any descendants of its children.  The \emph{height} of a block $b$ is its number of ancestors other than $b$, meaning that the height of $b_0$ is 0. 
By a \emph{chain} (ending in $b_h$), we mean a sequence of blocks $b_0 \ast b_1 \ast \dots \ast b_h$, such that $b_{h'+1}$ is a child of $b_{h'}$ for $h'<h$.\footnote{Throughout this paper, `$\ast$' denotes concatenation.} 


\subsection{Overview of the Snowman protocol}

To implement SMR, our approach is to run multiple instances of Snowflake$^+$. 
To keep things simple, consider first the task of reaching consensus on a block of height 1. Suppose that multiple children of $b_0$ are proposed over the course of the execution and that we must choose between them. To turn this decision problem into multiple binary decision problems,  we consider the hash value $H(b_1)$ of each proposed block $b_1$ of height 1, and then run one instance of Snowflake$^+$ to reach consensus on the first bit of the hash. Then we run  a second instance to reach consensus on the second bit of the hash, and so on. Working above a block of any height $h$, the same process is then used to finalize a block of height $h+1$. In this way, multiple instances of Snowflake$^+$ are used to reach consensus on a chain of hash values $H(b_0) \ast H(b_1) \ast \dots$. 

\vspace{0.2cm} 
This process would not be efficient if each round required a separate set of  correspondences for each instance of Snowflake$^+$, but this is not necessary. Just as in Snowflake$^+$, at the beginning of each round $s$, processor $p_i$ samples a single sequence $\langle p_{1,s},\dots p_{k,s} \rangle$ of $k$ processors. Since we now wish to reach consensus on a sequence of blocks, each processor $p_{j,s}$ in the sample is now requested to report its presently preferred chain, rather than a single bit value. The first bit of the corresponding hash sequence is then used by $p_i$ as the response of $p_{j,s}$ in a first instance of Snowflake$^+$. If this first bit agrees with $p_i$'s resulting value in that instance of Snowflake$^+$, then the second bit  is used as the value reported by $p_{j,s}$ in a second instance of Snowflake$^+$, and so on. 


\vspace{0.2cm} 
\noindent \textbf{A note on some simplifications that are made for the sake of clarity of presentation}.  When a processor $p_{j,s}$ is requested by $p_i$ to report its presently preferred chain (ending with $b$, say), we have $p_{j,s}$ simply send the given sequence of blocks. In reality, this would be very inefficient and the present implementation of Snowman deals with this by having $p_{j,s}$ send a hash of $b$ instead. This potentially causes some complexities, because $p_i$ may not have seen  $b$ (meaning that it does not necessarily know how to interpret the hash). This issue is easily dealt with, but it would be a distraction to go into the details here.

\vspace{0.2cm} 
\noindent \textbf{The variables, functions and procedures used by $p_i$}. The protocol instructions make use of the following variables and functions (as well as others whose use should be clear from the pseudocode):

\begin{itemize}
\item $b_0$: The genesis block.
\item $\mathtt{blocks}$: Stores blocks received by $p_i$ (and verified as valid). Initially it contains only $b_0$, and it is automatically updated over time to include any block included in any message received or sent by $p_i$.
\item $\mathtt{val}(\sigma)$: For each finite binary string $\sigma$, $\mathtt{val}(\sigma)$ records $p_i$'s presently preferred value for the next bit of the chain of hash values $H(b_0) \ast H(b_1) \ast \dots$, should the latter extend $\sigma$. 
\item $\mathtt{pref}$: The initial segment of the chain of hash values that $p_i$ presently prefers. We write $|\mathtt{pref}|$ to denote the length of this binary string.
\item $\mathtt{final}$: The initial segment of the chain of hash values that $p_i$ presently regards as final. 

\item $\mathtt{chain}(\sigma)$: If there exists a greatest $h\in \mathbb{N}$ such that $\sigma=H(b_0) \ast \dots \ast H(b_h) \ast \tau$ for a chain of blocks $b_0 \ast \dots \ast b_h$ all seen by $p_i$, and for some finite string $\tau$, then $\mathtt{chain}(\sigma):= b_0 \ast \dots \ast b_h$. Otherwise, $\mathtt{chain}(\sigma):=b_0$. 

\item $\mathtt{reduct}(\sigma)$: If there exists a greatest $h\in \mathbb{N}$ such that $\sigma=H(b_0) \ast \dots \ast H(b_h) \ast \tau$ for a chain of blocks $b_0 \ast \dots \ast b_h$ all seen by $p_i$, and for some finite string $\tau$, then $\mathtt{reduct}(\sigma):=H(b_0) \ast \dots \ast H(b_h)$. Otherwise, $\mathtt{reduct}(\sigma):=H(b_0)$. 

\item $\mathtt{last}(\sigma)$: If there exists a greatest $h\in \mathbb{N}$ such that $\sigma=H(b_0) \ast \dots \ast H(b_h) \ast \tau$ for a chain  $b_0 \ast \dots \ast b_h$ all seen by $p_i$, and for some finite string $\tau$, then $\mathtt{last}(\sigma):= b_h$. Otherwise, $\mathtt{last}(\sigma):=b_0$.
\item $H_B$: If $B=b_0\ast b_1 \ast \dots \ast b_h$ is a chain, then $H_B:=H(b_0) \ast H(b_1) \ast \dots H(b_h)$, and if not then $H_B$ is the empty string $\emptyset$.
   
\end{itemize}

The pseudocode is described in Algorithm 2. For strings $\sigma$ and $\tau$, we write $\sigma \subseteq \tau$ to denote that $\sigma$ is an initial segment of $\tau$. For the sake of simplicity, the pseudocode considers a fixed value for $\alpha_2$, but one could also incorporate approaches such as Error-Driven Snowflake$^+$, described in Section \ref{flex}.

\begin{algorithm} \label{pc:Snowman}
\caption{Snowman: The instructions for processor $p_i$}
\begin{algorithmic}[1]

    \State \textbf{Inputs} 
    \State $\Delta, k,\alpha_1,\alpha_2,\beta \in \mathbb{N}$  

    \State \textbf{Local values} 
    
   \State $\mathtt{val}(\sigma)$, initially undefined 
    \State $\mathtt{count}(\sigma)$, initially set to $0$ 
    \State $\mathtt{rpref}(j,s)$, initially undefined    \Comment{Records preferred chain of $p_{j,s}$}       

    \State $\mathtt{blocks}$, initially contains just $b_0$ \Comment{Automatically updated}

    \State $\mathtt{pref}$, initially set to $H(b_0)$

    \State $\mathtt{final}$, initially set to $H(b_0)$
     
    \State 

  \State \textbf{The instructions for round $s$, beginning at time $2\Delta s$:}

  \State \hspace{0.3cm} Form sample sequence $\langle p_{1,s},\dots p_{k,s} \rangle$;           \Comment Sample  with replacement

  \State \hspace{0.3cm} For $j\in [1,k]$, send $s$ to $p_{j,s}$;    \Comment Ask  $p_{j,s}$ for preferred chain

  \State 

   \State \hspace{0.3cm}  Wait $\Delta$;

     \State 

   \State \hspace{0.3cm} For each $j$ such that $p_i$ has received $s$ from $p_j$:

   \State \hspace{0.8cm} Send $(s,\mathtt{chain}(\mathtt{pref}))$ to $p_j$; \Comment{Report preferred chain to $p_j$}

     \State 

   \State \hspace{0.3cm}  Wait $\Delta$;

     \State 

   \State \hspace{0.3cm} For each $j\in [1,k]$:
   \State \hspace{0.8cm} \textbf{If} $p_i$ has received a first message $(s,B)$ from $p_{j,s}$ s.t.\ $B$ is a chain;
   \State \hspace{1.3cm}  Set $\mathtt{rpref}(j,s):=H_B$; \Comment{Record preferred chain of $p_{j,s}$}
   \State \hspace{0.8cm} \textbf{Else} set $\mathtt{rpref}(j,s):= H(b_0)$; 

  \State 

   \State \hspace{0.3cm} Set $\mathtt{pref}:=\mathtt{final}$, $\mathtt{end}:=0$; \Comment{Begin iteration to determine $\mathtt{pref}$ for round $s+1$} 

  \State \hspace{0.3cm} \textbf{While} $\mathtt{end}==0$ \textbf{do}: 

  \State \hspace{0.8cm} Set $E:= \{ b\in \mathtt{blocks}:\ b \text{ is a child of }\mathtt{last}(\mathtt{pref}) \text{ and }\mathtt{pref} \subseteq \mathtt{reduct}(\mathtt{pref})\ast H(b)\}$;

  \State \hspace{0.8cm} \textbf{If} $E$ is empty, set $\mathtt{end}:=1$;

 \State \hspace{0.8cm} \textbf{Else}:        \Comment{Carry out the next instance of Snowflake$^+$}

\State \hspace{1.3cm} \textbf{If} $\mathtt{val}(\mathtt{pref})$ is undefined:

\State \hspace{1.8cm} Let $b$ be the first block in $E$ enumerated into $\mathtt{blocks}$;  

\State \hspace{1.8cm} Set $\mathtt{val}(\mathtt{pref})$ to be the $(|\mathtt{pref}|+1)^{\text{th}}$ bit of $\mathtt{reduct}(\mathtt{pref})\ast H(b)$;

\State \hspace{1.3cm} \textbf{If} $|\{ j\in [1,k]: \mathtt{rpref}(j,s)\supseteq \mathtt{pref} \ast 1-\mathtt{val}(\mathtt{pref}) \}|\geq \alpha_1$: 

\State \hspace{1.8cm} Set $\mathtt{val}(\mathtt{pref}):=1-\mathtt{val}(\mathtt{pref})$; For all $\sigma \supseteq \mathtt{pref}$, set $\mathtt{count}(\sigma):=0$; 

  \State \hspace{1.3cm} \textbf{If} $|\{ j\in [1,k]: \mathtt{rpref}(j,s)\supseteq \mathtt{pref} \ast \mathtt{val}(\mathtt{pref}) \}|< \alpha_2$:
  
\State \hspace{1.8cm} For all $\sigma \supseteq \mathtt{pref}$, set $\mathtt{count}(\sigma):=0$;

 \State \hspace{1.3cm} \textbf{If} $|\{ j\in [1,k]: \mathtt{rpref}(j,s)\supseteq \mathtt{pref} \ast \mathtt{val}(\mathtt{pref}) \}|\geq \alpha_2$:
  
\State \hspace{1.8cm} Set $\mathtt{count}(\mathtt{pref}):=\mathtt{count}(\mathtt{pref})+1$;

 \State \hspace{1.3cm} \textbf{If} $\mathtt{count}(\mathtt{pref})\geq \beta$:
 
 \State \hspace{1.8cm} Set $\mathtt{final}:= \mathtt{pref} \ast \mathtt{val}(\mathtt{pref})$;

 \State \hspace{1.3cm} Set $\mathtt{pref}:=\mathtt{pref} \ast \mathtt{val}(\mathtt{pref})$;





\end{algorithmic}
\end{algorithm}

\section{Consistency analysis for Snowman} \label{Snowmansecurity}

We write $\mathtt{pref}_i$ and $\mathtt{final}_i$ to denote the values $\mathtt{pref}$ and $\mathtt{final}$ as locally defined for $p_i$. We say \emph{$p_i$ finalizes $\sigma$}, or $\sigma$ \emph{becomes final for $p_i$}, if there exists some round during which $\sigma \subseteq \mathtt{final}_i$.  
 We say $\sigma$ \emph{becomes final} if it becomes final for all correct processors. A block $b$ becomes final if there exists some chain $B=b_0 \ast \dots \ast b$ such that $H_B$ becomes final.  
 
\vspace{0.2cm} 
\noindent \textbf{The requirements}. A probabilistic protocol for SMR is required to satisfy the following properties, except with small  error probability: 

\noindent \emph{Liveness}: Unboundedly  many blocks become final.\footnote{To ensure that transactions are not censored, it is natural also to require the stronger condition that unboundedly  blocks \emph{produced by correct processors} become final. We initially consider the version of liveness stated above for the sake of simplicity, but describe how to deal with the stronger requirement in Section \ref{Avalanche2analysis}. In Section \ref{Avalanche2analysis}, we will also analyse the maximum time required to finalize new values. }    

\noindent \emph{Consistency}: Suppose $\sigma:=\mathtt{final}_i$ as defined at the beginning of round $s$ and that 
$\sigma':=\mathtt{final}_j$  as defined at the beginning of round $s'$. Then, whenever $p_i$ and $p_j$ are correct: 
\begin{itemize} 
\item[(i)] If $i=j$ and $s'\geq s$ then $\sigma \subseteq \sigma'$. 
\item[(ii)] Either $\sigma$ extends $\sigma'$, or $\sigma'$ extends $\sigma$. 
\end{itemize}

In this section, we show that Snowman satisfies consistency (except with small  error probability) for appropriate choices of the protocol parameters, and so long as $f<n/5$. As in Section \ref{Snowflakeanalysis}, for the sake of concreteness we give an analysis for $k=80$, $\alpha_1=41$, $\alpha_2=72$, and $\beta=12$, under the assumption that the population size $n\geq 500$. As before, we make the assumption that $f<n/5$ and $n\geq 500$ only so as to be able to give as simple a proof as possible: a more fine-grained analysis for smaller $n$ is the subject of future work. 
In Section \ref{Avalanche2} we will describe  how to augment Snowman with a module guaranteeing liveness, which is formally analysed in Section \ref{Avalanche2analysis}.

\vspace{0.2cm} 
\noindent \textbf{The proof of consistency}. It follows directly from the protocol instructions that (i) in the definition of consistency is satisfied. To see this, note that, initially, $\mathtt{pref}_i=\mathtt{final}_i=H(b_0)$. The values $\mathtt{pref}_i$ and $\mathtt{final}_i$ are not redefined during round $s$ prior to line 27, when we set $\mathtt{pref}_i:=\mathtt{final}_i$. If $\mathtt{pref}_i$ is subsequently redefined during round $s$, then we redefine it to be an extension of its previous value. If $\mathtt{final}_i$ is redefined during round $s$, then it is defined to be an extension of the present value of $\mathtt{pref}_i$. 

\vspace{0.2cm} To argue that (ii) in the definition of consistency is satisfied,  we again adopt the conventions described in Section \ref{model} concerning the treatment of small  error probabilities and suppose the protocol is run by at most 10,000 processors for at most 1000 years, executing at most 5 rounds per second.  We'll say a correct processor $p_i$ `samples $x$ values extending $\sigma$' in round $s$ if
$|\{ j: 1\leq j \leq k, \mathtt{rpref}(j,s)  \supseteq \sigma  \}|=x$, where $\mathtt{rpref}(j,s)$ is as locally defined for $p_i$ at the end of round $s$. 
We define $\sigma_s$ to be the longest string such that at least 75\% of correct processors have local $\mathtt{pref}$ values extending $\sigma_s$ at the beginning of round $s$, and such that no correct processor finalizes any value incompatible with $\sigma_s$ in any round $<s$.  We define $\sigma_s^*$ to be the longest string such that at least 75\% of correct processors have local $\mathtt{pref}$ values extending $\sigma_s^*$ at the beginning of round $s$. 

\vspace{0.2cm} 
The argument consists of four parts, similar to those described in Section \ref{Snowflakeanalysis}.

\vspace{0.2cm} 
\noindent \textbf{Part 1}. As in  Section \ref{Snowflakeanalysis}, a calculation for the binomial distribution shows that the probability a given correct processor has local $\mathtt{pref}$ value extending $\sigma_s$  at the beginning of round $s+1$ is  at least 0.9555, i.e.\ $\text{Bin}(80,0.8 \times 0.75,\geq41)>0.9555$. To see this, note that if $p_i$ samples at least 41 values extending $\sigma_s$ in round $s$, then, by the definition of $\sigma_s$,  it must set $\mathtt{pref}_i$ to be compatible with $\sigma_s $ in line 27 during round $s$. The \textbf{while} loop (lines 28--43) will then set $\mathtt{pref}_i$ to be an extension of $\sigma_s $ (possibly $\sigma_s$ itself). Assuming a population of at least 500, of which at least 80\%  are correct, another direct calculation for the binomial distribution shows that the probability that it fails to be the case that more than 5/6 of the correct processors have local $\mathtt{pref}$ values extending $\sigma_s$ at the beginning of round $s+1$ is upper bounded by $1.59\times 10^{-20}$,  i.e.\ $\text{Bin}(n,0.9555,\leq 5n/6)<1.59 \times 10^{-20}$ for $n\geq 400$.  If the protocol is executed for at most 1000 years, with at most 5 rounds executed per second, this means that less than $1.6\times 10^{11}$ rounds are executed. Applying the union bound, we conclude that the statement below holds, except with probability at most  $(1.6\times 10^{11}) \times (1.59\times 10^{-20})  < 3\times 10^{-9}$:  

\begin{enumerate} 
\item[$(\dagger_1)$] For every $s$, more than $5/6$ of correct processors have local $\mathtt{pref}$ values extending $\sigma_s$ at the beginning of round $s+1$. 
\end{enumerate}

\vspace{0.2cm} 
\noindent \textbf{Part 2}.  A  calculation for the binomial distribution shows that the probability that a given correct processor $p_i$ samples at least 72 values incompatible with $\sigma_s^*$  in round $s$ is upper bounded by $1.18 \times 10^{-20}$, i.e.\ $\text{Bin}(80,0.2+(0.8 \times 0.25),\geq 72)<1.18 \times 10^{-20}$. If the protocol is run by at most 10,000 processors for at most 1000 years, executing at most 5 rounds per second, then we can apply the union bound to deduce that the following statement holds, except with probability at most $(1.18\times 10^{-20})\times 10000 \times (1.6 \times 10^{11})<  2\times 10^{-5}$:

\begin{enumerate}
\item[$(\dagger_2)$] For every $s$, no correct processor samples at least 72 values incompatible with  $\sigma_s^*$ in round $s$.
\end{enumerate}

\vspace{0.2cm} 
\noindent \textbf{Part 3}.  Now consider a given processor $p$ and  the probability, $x$ say,  that there exists some $\sigma \not \subseteq \sigma_s^*$ such that $p$ samples 72 or more values in round $s$  extending  $\sigma$.  Calculations for the binomial distribution show that (independent of events prior to round $s$), this probability is less than 0.0131. To see this note that $x<x_0+x_1+x_2$, where: 
\begin{itemize} 
\item $x_0$ is the probability that $p$ samples 72 or more values in round $s$  extending $\sigma_s^*\ast 0$. 
\item $x_1$ is the probability that $p$ samples 72 or more values in round $s$  extending $\sigma_s^*\ast 1$. 
\item $x_2$ is the probability that $p$ samples 72 or more values in round $s$ that are incompatible with $\sigma_s^*$. 
\end{itemize}  
The calculation from Part 2 already shows that $x_2<  1.18 \times 10^{-20}$. To bound $x_0$ and $x_1$, suppose first that at least 50\% of correct processors have local $\mathtt{pref}$ values extending $\sigma_s^*\ast 0$ at the beginning of round $s$. In this case, $x_0$ is at most $\text{Bin}(80,(0.75\times 0.8)+0.2,\geq 72)<0.01309$, while $x_1$ is at most  $\text{Bin}(80,(0.5\times 0.8)+0.2,\geq 72)<3\times 10^{-9}$. If less than  50\% of correct processors have local $\mathtt{pref}$ values extending $\sigma_s^*\ast 0$ at the beginning of round $s$, then $x_0<3\times 10^{-9}$, while $x_1<0.01309$. Either way $x_0+x_1+x_2<0.0131$, as claimed. 
This calculation held irrespective of events prior to round $s$.  So, if we consider any 12 given consecutive rounds $[s,s+11]$ and any given correct processor $p_i$, the probability that, for every $s'\in [s,s+11]$,  $p_i$ samples at least 72 values in round $s'$ that are not extended by $\sigma_{s'}^*$  is upper bounded by  $0.0131^{12}<10^{-22}$. If at most 10,000 processors execute the protocol for at most 1000 years, executing at most 5 rounds per second, we can then apply the union bound to conclude that the following statement fails to hold with probability at most $10^{-22}\times 10000 \times (1.6 \times 10^{11})<  2\times 10^{-7}$:

\begin{enumerate}
\item[$(\dagger_3)$] If a correct processor finalizes some string  $\sigma$ in some round $s+11$, then, for at least one round $s'\in [s,s+11]$, $\sigma_{s'}^*$ extends $\sigma$.
\end{enumerate}

\vspace{0.2cm} 
\noindent \textbf{Part 4}. Now we put parts 1--3 together. From the union bound and the analysis above, we may conclude that $(\dagger_1)-(\dagger_3)$ all hold, except with probability at most 
$(3\times 10^{-9})+(2\times 10^{-5})+( 2\times 10^{-7})<3\times 10^{-5}$. So, suppose that $(\dagger_1)-(\dagger_3)$ all hold. Define $\sigma_{-1}=\sigma_{-1}^*=H(b_0)$. We show by induction on rounds $\geq 0$ that: (a) $\sigma_s=\sigma_s^*$; (b) $\sigma_s\supseteq \sigma_{s-1}$, and; (c) if correct $p_i$ finalizes some $\sigma$ in a round $<s$, then $\sigma \subseteq \sigma_s$. The induction hypothesis clearly holds for round 0. Suppose that it holds for round $s$. From $(\dagger_1)$, it follows that $\sigma_{s+1}^*\supseteq \sigma_s$. 
From $(\dagger_2)$, it follows that no correct processor finalizes a value incompatible with $\sigma_{s}$ in round $s$, meaning that $\sigma_{s+1}\supseteq \sigma_s$.  From $(\dagger_3)$, it follows that if correct $p_i$ finalizes some string $\sigma$ during round $s$, then there exists $s'\in [s-11,s]$ with $\sigma_{s'}=\sigma_{s'}^*$ and $\sigma_{s'}\supseteq \sigma$. By the induction hypothesis, $\sigma_s\supseteq \sigma_{s'}$. Since $\sigma_{s+1}\supseteq \sigma_{s}$, it follows that $\sigma_{s+1}\supseteq \sigma$. So, $\sigma_{s+1}=\sigma_{s+1}^*$ and any string finalized by correct $p_i$ in a round $<s+1$ is extended by $\sigma_{s+1}$.  This suffices to show that the induction hypothesis holds for round $s+1$. Consistency is therefore satisfied, except with small probability. 



\section{Frosty} \label{Avalanche2}

Recall that our next aim is to augment Snowman with a liveness module, allowing us to guarantee liveness in the case that $f<n/5$. 

\subsection{Overview of Frosty}

In what follows, we assume that all messages are signed by the processor sending the message. We also suppose that $f<n/5$. Recall that the local variable $\mathtt{pref}$ is a processor's presently preferred chain and that $\mathtt{final}$ is its presently finalized chain.  

\vspace{0.2cm} 
\noindent \textbf{The use of epochs}. As outlined in Section \ref{intro}, the basic idea is to run the Snowman protocol during standard operation, and to temporarily fall back to a standard `quorum-based' protocol in the event that a substantial adversary attacks liveness for Snowman. We therefore consider instructions that are divided into \emph{epochs}. In the first epoch (epoch 0), processors implement Snowman. In the event of a liveness attack, processors then enter epoch 1 and implement the quorum-based protocol to finalize the next block. Once this is achieved, they enter epoch 2 and revert to Snowman, and so on. Processors only enter each odd epoch and start implementing the quorum-based protocol if a liveness attack during the previous epoch forces them to do so. The approach taken is reminiscent of Jolteon and Ditto \cite{ditto}, in the sense that a view/epoch change mechanism is used to move between an optimistic and fallback path. 

\vspace{0.2cm} 
\noindent \textbf{Adding a decision condition}. In even epochs, and when a processor sees sufficiently many consecutive rounds during which its local value $\mathtt{final}$ remains unchanged, it will send a message to others indicating that it wishes to proceed to the next epoch. Before any correct processor $p_i$ enters the next epoch, however, it requires messages from at least 1/5  of all processors indicating that they wish to do the same. This is necessary to avoid the adversary being able to trigger a change of epoch at will, but produces a difficulty:  some correct processors may wish to enter the next epoch, but the  number who wish to do so may not be enough to trigger the epoch change. To avoid such a situation persisting for an extended duration, we introduce an extra decision condition. Processors now report their value $\mathtt{final}$ as well as their value $\mathtt{pref}$ when sampled. We consider an extra parameter $\alpha_3$: for our analysis here, we suppose $\alpha_3=48$ (since $48=\frac{3}{5}\cdot 80$). If $p_i$ sees two consecutive samples in which at least $\alpha_3$ processors report $\mathtt{final}$ values that all extend $\sigma$, then $p_i$ will regard $\sigma$ as final. For $k=80$, $\alpha_3=48$ and if $f<n/5$, the probability that at least 3/5 of $p_i$'s sample sequence in a given round are Byzantine is less than $10^{-14}$, so the probability that this happens in two consecutive rounds is small. Except with small  probability, the new decision rule therefore only causes $p_i$ to finalize $\sigma$ in the case that a correct processor has already finalized this value, meaning that it is safe for $p_i$ to do the same. Using this new decision rule, we will be able to argue below that epoch changes are triggered in a timely fashion: either the epoch change is triggered soon after any correct $p_i$ wishes to change epoch, or else sufficiently many correct processors do not wish to trigger the change that $p_i$ is quickly able to finalize new values. 

\vspace{0.2cm} 
\noindent \textbf{Epoch certificates}. While in even epoch $e$, and for a parameter $\gamma$ (chosen to taste),\footnote{In Section \ref{Avalanche2analysis}, we suppose $\gamma\geq 300$.} $p_i$ will send the (signed) message $(\text{stuck},e,\mathtt{final})$ to all others when it sees $\gamma$ consecutive rounds during which its local value $\mathtt{final}$ remains unchanged. This message indicates that $p_i$ wishes to enter epoch $e+1$ and is referred to as an `epoch $e+1$ message'. For any fixed $\sigma$, a set of messages of size at least $n/5$, each signed by a different processor and of the form $(\text{stuck},e,\sigma)$, is called an \emph{epoch certificate} (EC) for epoch $e+1$.\footnote{To ensure ECs are strings of constant bounded length (independent of $n$), one could use standard `threshold' cryptography techniques \cite{boneh2001short,shoup2000practical}, but we will not concern ourselves with such issues here.} When $p_i$ sees an EC for epoch $e+1$, it will send the EC to all others and enter epoch $e+1$. This ensures that when any correct processor enters epoch $e+1$, all others will do so within time $\Delta$.

\vspace{0.2cm} 
\noindent \textbf{Ensuring consistency between epochs}. We must ensure that the value finalized by the quorum-based protocol during an odd epoch $e+1$ extends all $\mathtt{final}$ values for correct processors. To achieve this, the rough idea is that we have processors send out their local $\mathtt{pref}$ values upon entering epoch $e+1$, and then use these values to extract a chain that it is safe for the quorum based protocol to build on. 
Upon entering epoch $e+1$, we therefore have $p_i$ send out the message $(\text{start},e+1,\mathtt{pref})$. 
 This message is referred to as a \emph{starting vote} for epoch $e+1$ and, for any string $\sigma$, we say that the starting vote $(\text{start},e+1,\mathtt{pref})$ extends $\sigma$ if $\sigma \subseteq \mathtt{pref}$.  By a \emph{starting certificate} (SC) for epoch $e+1$ we mean a set of at least $2n/3$ starting votes for epoch $e+1$, each signed by a different processor. If $S$ is an SC for epoch $e+1$, we set $\mathtt{Pref}^*(S)$ to be the longest $\sigma$ extended by more than half of the messages in $S$. The basic idea is that $\mathtt{Pref}^*(S)$ must extend all $\mathtt{final}$ values for correct processors, and that consistency will therefore be maintained if we have the quorum-based protocol finalize a value extending this string. 

To argue that this is indeed the case, recall the proof described in Section \ref{Snowmansecurity} (and recall that $f<n/5$). We argued there that, if any correct processor $p_i$ finalizes $\sigma$ in a given round, then (except with small  error probability),  more than 5/6 of the honest processors must have local $\mathtt{pref}$ values that extend $\sigma$ by the end of that round, and that this will also be the case in all subsequent rounds. This might seem to ensure that $\mathtt{Pref}^*(S)$ will extend $\sigma$: since $\frac{5}{6}\cdot \frac{4}{5}=\frac{2}{3}$, and since $S$ contains at least $2n/3$ starting votes, it is tempting to infer that more than half the votes in $S$ must extend $\sigma$.  A complexity here, however,  is that this reasoning only applies if all $\mathtt{Pref}$ values are reported \emph{in the same round}. We can't (easily) ensure that all correct processors enter $e+1$ epoch in the same round, meaning that some correct processors may send their $\mathtt{Pref}$ values in one round, while others send them in the next round. 
 To deal with this, we increase the $\beta$ parameter from 12 to 14. This ensures (except with small  error probability) that, when a correct processor $p_i$ finalizes $\sigma$, more than 11/12 of correct processors have local $\mathtt{pref}$ values that extend $\sigma$ by the end of the previous round, and that this is also true in all subsequent rounds.  If $s$ and $s+1$ are two consecutive rounds after $p_i$ finalizes $\sigma$, and if we partition the correct processors arbitrarily so that some report their $\mathtt{pref}$  value in round $s$, while the rest do so in round $s+1$, then at least 5/6 of the correct processors must report values extending $\sigma$. 

\vspace{0.2cm} 
\noindent \textbf{The choice of quorum-based protocol}. While any of the standard quorum-based protocols could be implemented during odd epochs, for the sake of simplicity we give an exposition that implements a form of Tendermint, 
and we assume familiarity with that protocol in what follows. Let $f^*$ be the greatest integer less than $n/3$. Recall that the instructions for Tendermint are divided into rounds (sometimes called `views'). Within
each round, there are two stages of voting, each of which is an
opportunity for processors to vote on a block proposed by the
\emph{leader} of the round. The first stage of voting may establish
a \emph{stage 1 quorum certificate} (QC) for the proposed block, which is a set of stage 1 votes from $n-f^*$ distinct processors. In this event, the 
second stage may establish a \emph{stage 2} QC for the block. The protocol also implements a \emph{locking} mechanism. Processor $p_i$ maintains a value $\mathtt{Q}^+$. When they cast a stage 2 vote during round $s$, meaning that they have seen some $Q$ which is a stage 1 QC for the proposal they are voting on, they set $Q^+:=Q$.

In our version of Tendermint, each leader will make a \emph{proposal} $P$, and other processors will then vote on the proposal (so our `proposals' play the role of blocks in Tendermint). 

%
%
%
%
%
%
%
%
%
%
%
%
%
%
%
%
%
%
%
%

\subsection{Further terminology}


We consider the following new variables and other definitions (in addition to those used in previous sections). 

\vspace{0.1cm} 
\noindent $f^*$: The greatest integer less than $n/3$.

\vspace{0.1cm} 
\noindent $\mathtt{e}$: The epoch in which $p_i$ is presently participating (initially 0). 

\vspace{0.1cm} 
\noindent $\mathtt{stuckcount}$: Counts the number of consecutive rounds with $\mathtt{final}$ unchanged.

\vspace{0.1cm} 
\noindent $\mathtt{ready}(e)$: Indicates whether we have already initialized values for epoch $e$. Initially, $\mathtt{ready}(e)=0$. Processor $p_i$ sets $\mathtt{ready}(e):=1$ upon entering epoch $e$ after initializing values so that it is ready to start executing instructions for the epoch. 

\vspace{0.1cm} 
\noindent $\mathtt{Init}(e)$: This process is run at the beginning of even epoch $e$, and performs the following: Set $\mathtt{pref}:=\mathtt{final}$, $\mathtt{stuckcount}:=0$, and for all $\sigma$ set $\mathtt{count}(\sigma):=0$, $\mathtt{primed}(\sigma):=0$, and make $\mathtt{val}(\sigma)$ undefined. 

\vspace{0.1cm} 
\noindent $\mathtt{M}$: The set of all messages so far received by $p_i$.

\vspace{0.1cm} 
\noindent $\mathtt{lead}(s)$: The leader of round $s$ while in an odd epoch. We set $\mathtt{lead}(s)=p_j$, where $j= s \text{ mod }n$.

\vspace{0.1cm} 
\noindent $\mathtt{primed}(\sigma)$: Used to help implement the new decision rule. This value is initially 0, and is set to 1 when $p_i$
sees sufficiently many sampled $\mathtt{final}$ values extending $\sigma$. In the next round, $p_i$ either finalizes $\sigma$ (if the same holds again), or else resets $\mathtt{primed}(\sigma)$ to 0.  

\vspace{0.1cm} 
\noindent \textbf{Starting votes}: A starting vote for epoch $e$ is a message of the form 
$(\text{start},e,\sigma)$ for some $\sigma$. 
For any string $\sigma'$, we say that the starting vote $(\text{start},e,\sigma)$ extends $\sigma'$ if $\sigma' \subseteq \sigma$. 

\vspace{0.1cm} 
\noindent \textbf{Starting certificates}: A starting certificate for epoch $e$ is  set of at least $2n/3$ starting votes for epoch $e$, each signed by a different processor.

\vspace{0.1cm} 
\noindent $\mathtt{Pref}^*(S)$: If $S$ is a starting certificate (SC) for epoch $e$, we set $\mathtt{Pref}^*(S)$ to be the longest $\sigma$ extended by more than half of the messages in $S$.

\vspace{0.2cm} 
\noindent \textbf{Votes}.  A vote $V$ (for a proposal) is entirely specified by the following values: 

 \vspace{0.1cm} 
\noindent $\mathtt{P}(V)$: The proposal for which  $V$ is a vote. 

 \vspace{0.1cm} 
\noindent $\mathtt{st}(V)$: The stage of the vote (1 or 2).

\vspace{0.2cm} 
\noindent \textbf{The empty proposal}. We call $\emptyset$ the \emph{empty proposal}, and also let $\emptyset$ be a stage 1 QC for the empty proposal. 
We set $\mathtt{r}(\emptyset):=0$. The empty proposal is \emph{$M$-valid} for any set of messages $M$. 

\vspace{0.2cm} 
\noindent \textbf{Proposals}. A proposal $P$ other than the empty proposal is entirely specified by the following values: 

\noindent $\mathtt{r}(P)$: The round corresponding to the proposal. 

\noindent $\mathtt{e}(P)$: The epoch corresponding to the proposal. 

\noindent $\mathtt{par}(P)$: A proposal which is called the \emph{parent} of $\mathtt{P}$.\footnote{We adopt similar terminology for proposals and blocks:  If $P'$ is the parent of $P$, then $P$ is referred to as a \emph{child} of $P$. In this case, the ancestors of $P$ are $P$ and any ancestors of $P'$.  The descendants of any proposal $P$ are $P$ and any descendants of its children.} 

\noindent $\mathtt{QCprev}(P)$: A stage 1 QC for $\mathtt{par}(P)$. 

\noindent $\mathtt{final}(P)$: The value that $P$ attempts to finalize.

\noindent $\mathtt{SC}(P)$: A starting certificate justifying the proposed value for finalization. 

\vspace{0.2cm} 
\noindent \textbf{$M$-valid proposals}. Let $M$ be a set of messages. A proposal $P$ other than the empty proposal is $M$-valid if it satisfies all of the following: 
\begin{itemize} 
\item $P\in M$.
\item $P$ has the empty proposal as an ancestor.
\item $\mathtt{par}(P)$ is $M$-valid. 
\item If $\mathtt{par}(P)$ is not the empty proposal, then $\mathtt{e}(P)= \mathtt{e}(\mathtt{par}(P))$. 
\item If $\mathtt{par}(P)$ is not the empty proposal, then $\mathtt{final}(P)=\mathtt{final}(\mathtt{par}(P))$.
\item $\mathtt{QCprev}(P)$ is a stage 1 QC for $\mathtt{par}(P)$. 
\item $\mathtt{SC}(P)$ is a starting certificate for epoch $\mathtt{e}(P)$. 
\item $\mathtt{final}(P)$ extends $\mathtt{Pref}^*(\mathtt{SC}(P))$.
\end{itemize}

\vspace{0.2cm} 
\noindent \textbf{An $\mathtt{M}$-valid proposal for round $s$}. Let $\mathtt{M}$, $\mathtt{blocks}$ and $\mathtt{e}$ be as locally defined for $p_i$. While in round $s$, at time $3s\Delta +\Delta$, $p_i$ will regard the proposal $P$ as an $\mathtt{M}$-valid proposal for round $s$ if all of the following are satisfied: 
\begin{itemize} 
\item $P$ is $\mathtt{M}$-valid. 
\item $\mathtt{r}(P)=s$ and $P$ is signed by $\mathtt{lead}(s)$. 
\item $\mathtt{e}(P)=\mathtt{e}$. 
\item There exists a chain $B=b_0\ast \dots \ast b_h$ such that $\mathtt{final}(P)=H_B$ and, for $j\in [1,h]$, $b_j\in \mathtt{blocks}$. 
\end{itemize} 

\vspace{0.2cm} 
\noindent $M$-\textbf{confirmed proposals}. For any set of messages $M$, a proposal $P$ is $M$-confirmed if a descendant $P'$ of $P$ (possibly $P$ itself) is $M$-valid and $M$ contains a stage 2 QC for $P'$. 

\vspace{0.2cm} 
\noindent \textbf{Epoch certificates}.  For any fixed $\sigma$, a set of messages of size at least $n/5$, each signed by a different processor and of the form $(\text{stuck},e,\sigma)$, is called an \emph{epoch certificate} (EC) for epoch $e+1$.

\vspace{0.2cm} 
\noindent \textbf{Quorum certificates}. If $P$ is any proposal other than the empty proposal, then, by a stage $x$ QC for $P$, we mean a set of votes $Q$ of size $n-f^*$, each signed by a different processor, and such that $\mathtt{P}(V)=P$ and $\mathtt{st}(V)=x$ for each $V\in Q $. If $Q$ is a (stage 1 or 2) QC for $P$, then we define $\mathtt{r}(Q):=\mathtt{r}(P)$.

\vspace{0.2cm} 
\noindent \textbf{The procedure $\mathtt{MakeProposal}$}. If $p_i=\mathtt{lead}(s)$, then this procedure is used by $p_i$ while in odd epochs to send an appropriate proposal to all: 
\begin{itemize} 
\item If $p_i$ has not seen an SC for epoch $e$, then do not send any proposal. Otherwise, let $S$ be such an SC and proceed as follows. 
\item Let $s'$ be the greatest such that $\mathtt{M}$ contains an $\mathtt{M}$-valid proposal $P'$ with $\mathtt{r}(P')=s'$, $\mathtt{e}(P')=\mathtt{e}$ if $s'>0$, and such that $\mathtt{M}$  also contains $Q$ which is a stage 1 QC for $P'$.
\item Set  $\mathtt{r}(P):=s$, $\mathtt{e}(P):=\mathtt{e}$, $\mathtt{par}(P):=P'$, $\mathtt{QCprev}(P):=Q$. 
\item If $s'=0$, i.e.\ if $P'$ is the empty proposal, then proceed as follows. Set $\mathtt{SC}(P):=S$. Let $B=b_0\ast \dots \ast b_h$ be a chain such that $H_B$ extends $\mathtt{Pref}^*(S)$ and, for $j\in [1,h]$, $b_j\in \mathtt{blocks}$ (if there exists no such $B$ then do not send a proposal). Set $\mathtt{final}(P):= H_B$.
\item If $s'>0$, then set $\mathtt{SC}(P):=\mathtt{SC}(P')$ and $\mathtt{final}(P):=\mathtt{final}(P')$. 
\item Send $P$ to all. 
\end{itemize}

\vspace{0.2cm} 
\noindent \textbf{Conventions regarding the gossiping of blocks, proposals and QCs while in an odd epoch}. While in odd epochs, we suppose that correct processors automatically gossip received blocks, proposals and QCs for proposals. This means that if $p_i$ is correct and sees a QC (for example) at time $t$, then all correct processors see that QC by time $t+\Delta$. When $p_i$ sends a message to all, it is also convenient to assume $p_i$ regards that message as received (by $p_i$) at the next timeslot. 

\vspace{0.2cm} 
\noindent For ease of reference,  inputs and local variables are listed in the table below.  The pseudocode appears in Algorithms 3 and 4.

\vspace{0.3cm} 
\begin{tabular}{l}

\hline

\noindent Frosty: The inputs and local values for processor $p_i$ \\

\hline

\noindent \textbf{Inputs} \\

\noindent $\Delta, k,\alpha_1,\alpha_2,\alpha_3,\beta,\gamma \in \mathbb{N}$  \\


\noindent \textbf{Local values} \\ 

\noindent $\mathtt{val}(\sigma)$, initially undefined. \\

\noindent $\mathtt{count}(\sigma)$, initially set to $0$ \\

\noindent $\mathtt{primed}(\sigma)$, initially set to $0$ \\

\noindent $\mathtt{stuckcount}$, initially set to $0$ \\

\noindent $\mathtt{rpref}(j,s)$, initially undefined   \\

\noindent $\mathtt{final}(j,s)$, initially undefined \\

\noindent $\mathtt{blocks}$, initially contains just $b_0$ \\

\noindent $\mathtt{pref}$, initially set to $H(b_0)$ \\

\noindent $\mathtt{final}$, initially set to $H(b_0)$ \\

\noindent $\mathtt{e}$, initially set to 0 \\

\noindent $\mathtt{ready}(e)$, initially set to 0 for all $e\in \mathbb{N}_{\geq 0}$ \\

\noindent $\mathtt{Q}^+$, initially set to $\emptyset$ \\

\noindent $\mathtt{M}$, initially contains just $b_0$ \\

\noindent $\mathtt{P}$, initially undefined \\

\hline

\end{tabular}

\begin{algorithm} \label{pc:Ava1.5}
\caption{Frosty: The instructions for processor $p_i$ $\textbf{while}\  \mathtt{e}$ is even}
\begin{algorithmic}[1]

\State At every $t$ \textbf{if} $\mathtt{ready}(\mathtt{e})==0$ \textbf{then} $\mathtt{Init}(e)$, set $\mathtt{ready}(e):=1$;       \Comment{Initialise values for epoch $\mathtt{e}$}

    \State 

  \State At $t=3\Delta s$ \textbf{if} $\mathtt{ready}(\mathtt{e})==1$:      \Comment{For any $s\in \mathbb{N}_{\geq 1}$}

  \State \hspace{0.3cm} Form sample sequence $\langle p_{1,s},\dots p_{k,s} \rangle$;           \Comment Sample  with replacement

  \State \hspace{0.3cm} For $j\in [1,k]$, send $(s,\mathtt{e})$ to $p_{j,s}$;    \Comment Ask  $p_{j,s}$ for preferred chain

    \State \hspace{0.3cm} \textbf{If} $\mathtt{M}$ contains an EC for epoch $\mathtt{e}+1$: send the EC to all, set $\mathtt{e}:=\mathtt{e}+1$;   \Comment{Enter next epoch}

     \State 

      \State At $t=3\Delta s+\Delta$ \textbf{if} $\mathtt{ready}(\mathtt{e})==1$:

   \State \hspace{0.3cm} For each $j$ such that $p_i$ has received $(s,\mathtt{e})$ from $p_j$:

   \State \hspace{0.6cm} Send $(s,\mathtt{e},\mathtt{chain}(\mathtt{pref}),\mathtt{chain}(\mathtt{final}))$ to $p_j$; \Comment{Report present values to $p_j$}

    \State \hspace{0.3cm} \textbf{If} $\mathtt{M}$ contains an EC for epoch $\mathtt{e}+1$: send the EC to all, set $\mathtt{e}:=\mathtt{e}+1$;   \Comment{Enter next epoch}

     \State 

        \State At $t=3\Delta s+2\Delta$ \textbf{if} $\mathtt{ready}(\mathtt{e})==1$:

   \State \hspace{0.3cm} Form sample sequence $\langle p_{1,s},\dots p_{k,s} \rangle$ if not already formed.
   \State \hspace{0.3cm}  For each $j\in [1,k]$:
   \State \hspace{0.6cm} \textbf{If} $p_i$ has received a first message $(s,\mathtt{e},B_1,B_2)$ from $p_{j,s}$ s.t.\ $B_1,B_2$ are chains;
   \State \hspace{0.9cm}  Set $\mathtt{rpref}(j,s):=H_{B_1}$, $\mathtt{final}(j,s):=H_{B_2}$; \Comment{Record values from $p_{j,s}$}
   \State \hspace{0.6cm} \textbf{Else} set $\mathtt{rpref}(j,s):= H(b_0)$,  $\mathtt{final}(j,s):= H(b_0)$; 
        
    \State \hspace{0.3cm} \textbf{If} $\mathtt{M}$ contains an EC for epoch $\mathtt{e}+1$: send the EC to all, set $\mathtt{e}:=\mathtt{e}+1$;   \Comment{Enter next epoch}

     \State    \State At $t=3\Delta s+2\Delta$ \textbf{if} $\mathtt{ready}(\mathtt{e})==1$:   \Comment{Execute instructions above first and re-check $\mathtt{ready}(\mathtt{e})$}

 \State \hspace{0.3cm} Set $E^*:= \{ b\in \mathtt{block}:\ b \text{ is a child of }\mathtt{last}(\mathtt{final}) \}$; 

 \State \hspace{0.3cm} Set $\mathtt{pref}:=\mathtt{final}$, $\mathtt{end}:=0$. If $E^*$ is non-empty: $\mathtt{stuckcount}++$;

  \State \hspace{0.3cm} \textbf{While} $\mathtt{end}==0$ \textbf{do}:  \Comment{Iteration to determine $\mathtt{pref}$ for round $s+1$}

  \State \hspace{0.6cm} Set $E:= \{ b\in \mathtt{block}:\ b \text{ is a child of }\mathtt{last}(\mathtt{pref}) \text{ and }\mathtt{pref} \subseteq \mathtt{reduct}(\mathtt{pref})\ast H(b)\}$;

  \State \hspace{0.6cm} \textbf{If} $E$ is empty, set $\mathtt{end}:=1$;

 \State \hspace{0.6cm} \textbf{Else}:        \Comment{Carry out the next instance of Snowflake$^+$}

\State \hspace{0.9cm} \textbf{If} $\mathtt{val}(\mathtt{pref})$ is undefined:

\State \hspace{1.2cm} Let $b$ be the first block in $E$ enumerated into $\mathtt{block}$;  

\State \hspace{1.2cm} Set $\mathtt{val}(\mathtt{pref})$ to be the $(|\mathtt{pref}|+1)^{\text{th}}$ bit of $\mathtt{reduct}(\mathtt{pref})\ast H(b)$;

\State \hspace{0.9cm} \textbf{If} $|\{ j\in [1,k]: \mathtt{rpref}(j,s)\supseteq \mathtt{pref} \ast 1-\mathtt{val}(\mathtt{pref}) \}|\geq \alpha_1$: 

\State \hspace{1.2cm} Set $\mathtt{val}(\mathtt{pref}):=1-\mathtt{val}(\mathtt{pref})$;  For all $\sigma \supseteq \mathtt{pref}$, set $\mathtt{count}(\sigma):=0$;

  \State \hspace{0.9cm} \textbf{If} $|\{ j\in [1,k]: \mathtt{rpref}(j,s)\supseteq \mathtt{pref} \ast \mathtt{val}(\mathtt{pref}) \}|< \alpha_2$: 
  
  \State \hspace{1.2cm} For all $\sigma \supseteq \mathtt{pref}$, set $\mathtt{count}(\sigma):=0$;

 \State \hspace{0.9cm} \textbf{If} $|\{ j\in [1,k]: \mathtt{rpref}(j,s)\supseteq \mathtt{pref} \ast \mathtt{val}(\mathtt{pref}) \}|\geq \alpha_2$:
   $\mathtt{count}(\mathtt{pref})++$;

 \State \hspace{0.9cm} \textbf{If} $\mathtt{count}(\mathtt{pref})\geq \beta$:
 
 \State \hspace{1.2cm} Set $\mathtt{final}:= \mathtt{pref} \ast \mathtt{val}(\mathtt{pref})$, $\mathtt{stuckcount}:=0$; \Comment{Finalize and reset $\mathtt{stuckcount}$}

  \State \hspace{0.9cm} \textbf{If} $\mathtt{count}(\mathtt{pref})< \beta$ \textbf{then} for $x\in \{ 0,1 \}$ \textbf{do}: \Comment{New decision rule}

  \State \hspace{1.2cm} \textbf{If} $|\{ j\in [1,k]: \mathtt{final}(j,s)\supseteq \mathtt{pref} \ast x \}|\geq \alpha_3$: 

   \State \hspace{1.5cm} \textbf{If} $\mathtt{primed}(\mathtt{pref} \ast x)==1$; \Comment{previous round primed $\mathtt{pref}\ast x$ for finalization }

     \State \hspace{1.8cm}  Set  $\mathtt{final}:= \mathtt{pref} \ast x$,  $\mathtt{stuckcount}:=0$; \Comment{Finalize and reset $\mathtt{stuckcount}$}

     \State \hspace{1.5cm} \textbf{Else} set $\mathtt{primed}(\mathtt{pref}\ast x):=1$;  \Comment{Prime $\mathtt{pref}\ast x$ for finalization }

       \State \hspace{1.2cm} \textbf{Else} set $\mathtt{primed}(\mathtt{pref}\ast x):=0$;

 \State \hspace{0.9cm} Set $\mathtt{pref}:=\mathtt{pref} \ast \mathtt{val}(\mathtt{pref})$;

   \State \hspace{0.3cm} \textbf{If} $\mathtt{stuckcount}\geq \gamma$  \textbf{then} send $(\text{stuck},\mathtt{e},\mathtt{final})$ to all;   \Comment{After completing \textbf{while} loop}

\end{algorithmic}
\end{algorithm}

\begin{algorithm} \label{pc:Ava1.5odde}
\caption{Frosty: The instructions for processor $p_i$ $\textbf{while}\  \mathtt{e}$ is odd}
\begin{algorithmic}[1]

\State At every $t$ \textbf{if} $\mathtt{ready}(\mathtt{e})==0$ \textbf{then}:

  \State \hspace{0.3cm} Send $(\text{start},\mathtt{e},\mathtt{pref})$ to all; \Comment{Send starting vote}

  \State \hspace{0.3cm} Set $\mathtt{Q}^+:=\emptyset$;   \Comment{Initialize $\mathtt{Q}^+$}

  \State \hspace{0.3cm} Set $\mathtt{ready}(e):=1$;

    \State 

    \State At every $t$, \textbf{if} there exists proposal $P\in \mathtt{M}$ with $\mathtt{e}(P)==\mathtt{e}$ which is $\mathtt{M}$-confirmed \textbf{then}:

     \State \hspace{0.3cm} Set $\mathtt{final}:= \mathtt{final}(P)$;   \Comment{Finalize next block}

     \State \hspace{0.3cm} Set $\mathtt{e}:=\mathtt{e}+1$;   \Comment{Enter next epoch. Others will do so within $\Delta$ (due to gossiping)}

    \State 

  \State At $t=3\Delta s$ \textbf{if} $\mathtt{lead}(s)==i$:     \Comment{For any $s\in \mathbb{N}_{\geq 1}$}

    \State \hspace{0.3cm} $\mathtt{MakeProposal}$;

      \State

     \State At $t=3\Delta s +\Delta$:

      \State \hspace{0.3cm} Set $\mathtt{P}$ to be undefined. 
     
    \State \hspace{0.3cm}   \textbf{If} $p_i$ has received a first $\mathtt{M}$-valid proposal $P$ for round $s$ \textbf{and} $\mathtt{r}(\mathtt{QCprev}(P))\geq \mathtt{r}(\mathtt{Q}^+)$:

     \State \hspace{0.6cm} Set $\mathtt{P}:=P$;

      \State \hspace{0.6cm} Send vote $V$ to all, with $\mathtt{P}(V):=\mathtt{P}$, $\mathtt{st}(V):=1$; \Comment{Send stage 1 vote}

     \State 

       \State At $t=3\Delta s +2\Delta$:

          \State \hspace{0.3cm}   \textbf{If} $\mathtt{P}$ is defined and $p_i$ has received a first $Q$ which is a stage 1 QC for $\mathtt{P}$:

           \State \hspace{0.6cm}  Set $\mathtt{Q}^+:=Q$;  \Comment{Set lock}

             \State \hspace{0.6cm} Send vote $V$ to all, with $\mathtt{P}(V):=\mathtt{P}$, $\mathtt{st}(V):=2$; \Comment{Send stage 2 vote}

\end{algorithmic}
\end{algorithm}

\section{Frosty: Consistency and liveness analysis} \label{Avalanche2analysis}

We give an analysis for the case that $k=80$, $\alpha_1=41$, $\alpha_2=72$, $\alpha_3=48$, $\beta=14$, $\gamma\geq 300$, and under the assumption that $n\geq 500$ and $f<n/5$. As before, we make the assumption that $n\geq 500$ only so as to be able to give as simple a proof as possible: a more fine-grained analysis for smaller $n$ is the subject of future work. 

\subsection{The proof of consistency} \label{consis}

 Section \ref{Snowmansecurity} already established that Snowman satisfies consistency (except with small  error probability). To establish consistency for Frosty, we must show that, if an odd epoch $e$ finalizes any value, then it finalizes a single value extending any values finalized by correct processors during previous epochs. Before considering the instructions during odd epochs, however, there are three new complexities with respect to the instructions during an even epoch $e$, which we must check cannot lead to a consistency violation during the same epoch: 
 \begin{itemize} 
 \item[(i)] The new decision rule. 
 \item[(ii)] Players may not enter epoch $e+1$ entirely simultaneously. 
 \item[(iii)] Players may not enter epoch $e$ entirely simultaneously. 
   \end{itemize}

\vspace{0.2cm}
\noindent \textbf{Dealing with (i)}. Suppose $p_i$ is correct. A calculation for the binomial distribution shows that the probability that at least 3/5 of $p_i$'s sample sequence in a given round are Byzantine is less than $10^{-14}$. The probability that this happens in two given consecutive rounds is therefore less than $10^{-28}$. We may therefore condition on the following event (letting $\mathtt{final}(j,s)$ and $\mathtt{final}(j,s+1)$  be as locally defined for $p_i$ at the end of rounds $s$ and $s+1$ respectively): 
\begin{enumerate}
    \item[$(\diamond_0)$:] If there exists $s$ and $\sigma$ such that $|\{ j\in [1,k]: \mathtt{final}(j,s)\supseteq \sigma \}|\geq \alpha_3$ and $|\{ j\in [1,k]: \mathtt{final}(j,s+1)\supseteq \sigma \}|\geq \alpha_3$, then some correct processor has already finalized a value extending $\sigma$ by the end of round $s$. 
\end{enumerate}
Conditioned on $(\diamond_0)$, it is not possible for the new decision rule to cause a first consistency violation. 

\vspace{0.2cm}
\noindent \textbf{Dealing with (ii)}. If any correct processor enters epoch $e+1$ at $t$, then they send an EC for epoch $e+1$ to all. All correct processors therefore enter epoch $e+1$ by $t+\Delta$. In particular, this means that if correct $p_i$ is in epoch $e$ at time $3s\Delta +2\Delta$ (when considering values reported during round $s$), no correct processor will have entered epoch $e+1$ prior to reporting their values during round $s$ (lines 8--11 of the pseudocode). The distribution on reported values is thus unaffected by the fact that some correct processors may have already entered epoch $e+1$ at that point. 

\vspace{0.2cm}
\noindent \textbf{Dealing with (iii)}. If $e>0$ is even and some correct processor enters epoch $e$ at $t$ because there exists $P\in \mathtt{M}$ with $\mathtt{e}(P)=e-1$ which is $\mathtt{M}$-confirmed, then (due to the gossiping of blocks and QCs) all correct processors will enter epoch $e$ by $t+\Delta$. The arguments of Section \ref{Snowmansecurity} are unaffected if we allow that some correct processors only begin executing instructions midway through the first round of an epoch (this just means that some correct processors might not report values or ask for values in the first round). 

\vspace{0.2cm} Let $\mathtt{M}$ be as locally defined for correct $p_i$ and say that a proposal $P$ \emph{becomes confirmed for $p_i$} if there is some timeslot at which $P$ is $\mathtt{M}$-confirmed. Suppose $e$ is odd and that no consistency violation occurs prior to the first point at which any correct processor enters epoch $e$.  To complete the proof of consistency, it suffices to establish the two following claims: 

\vspace{0.1cm} 
\noindent \emph{Claim 1}. If the proposal $P$ with $\mathtt{e}(P)=e$ becomes confirmed for correct $p_i$, then $\mathtt{final}(P)$ extends any values finalized by correct processors during previous epochs. 

\vspace{0.1cm} 
\noindent \emph{Claim 2}. If $P$ and $P'$ are proposals with  $\mathtt{e}(P)=\mathtt{e}(P')=e$, and if $P$ becomes confirmed for correct $p_i$ and $P'$ becomes confirmed for correct $p_j$, then $\mathtt{final}(P)=\mathtt{final}(P')$. 

\vspace{0.2cm} 
\noindent \textbf{Establishing Claim 1}. If the proposal $P$ becomes confirmed for correct $p_i$, then some correct processors must produce votes for the proposal, which implies that: 
\begin{itemize}
\item $\mathtt{SC}(P)$ is a starting certificate for epoch $e$. 
\item $\mathtt{final}(P)$ extends $\mathtt{Pref}^*(\mathtt{SC}(P))$.
\end{itemize} 
We argued in Section \ref{Snowmansecurity} that, when $\beta=12$, if some correct processor is the first to finalize the value $\sigma$ and does so in some round $s+11$, say, then at least one round $s'\in [s,s+11]$ must satisfy the condition that at least 75\% of correct processors have local $\mathtt{pref}$ values extending $\sigma$ in round $s'$ and that at least 5/6 of the correct processors must have local $\mathtt{pref}$ values extending $\sigma$ (by the end of round $s'$ and) in all rounds $>s'$. Now consider Frosty for the case that $\beta=14$. Suppose that some correct processor is the first to finalize the value $\sigma$ while in  epoch $e-1$  and does so in some round $s+13$ (the case that no processor finalizes any new values while in epoch $e-1$ is similar).  Then, by the same reasoning as in Section \ref{Snowmansecurity}, at least one round $s'\in [s,s+11]$ must satisfy the condition that at least 5/6 of correct processors have local $\mathtt{pref}$ values extending $\sigma$ at the end of round $s'$. Let $s''$ be the greatest round such that some correct processor completes round $s''$ before entering epoch $e$.  A calculation for the binomial then shows that, except with small  error probability, at the end of each round in $[s+12,s'']$ (and even if some correct processors have already moved to epoch $e$ during round $s''$), at least 11/12 of the correct processors must have local $\mathtt{pref}$ values extending $\sigma$. Note that the local $\mathtt{pref}$ values reported by correct processors in the starting certificate $\mathtt{SC}(P)$ are either those defined at the end of round $s''$ or round $s''-1$. We conclude that at least 5/6 of correct processors must send starting votes of the form $(\text{start},e,\mathtt{pref})$ such that $\mathtt{pref}$ extends $\sigma$. Since $\mathtt{SC}(P)$ contains at least $2n/3$ starting votes,  more than half the votes in $\mathtt{SC}(P)$ must extend $\sigma$, so that $\mathtt{final}(P)$ extends $\sigma$, as required. 

\vspace{0.2cm} 
\noindent \textbf{Establishing Claim 2}. Towards a contradiction, suppose there exists some least $s$ and some least $s'\geq s$ such that: 
\begin{itemize}
    \item Some proposal $P$ with $\mathtt{e}(P)=e$ and $\mathtt{r}(P)=s$ receives stage 1 and 2 QCs, $Q_1$ and $Q_2$ respectively;
    \item Some proposal $P'$ with $\mathtt{e}(P')=e$ and $\mathtt{r}(P)=s'$ receives a stage 1 QC, $Q_1'$; 
    \item $\mathtt{final}(P)\neq \mathtt{final}(P')$. 
\end{itemize}
Suppose first that $s=s'$. Then, since each QC contains votes from at least $n-f^*$ distinct processors, some correct processor must produce votes in both $Q_1$ and $Q_1'$. This gives an immediate contradiction, because correct processors do not produce more than one stage 1 vote in any single round.

So, suppose that $s'>s$. In this case, some correct processor $p_i$ must produce votes in both $Q_2$ and $Q_1'$. Since $\mathtt{final}(P)\neq\mathtt{final}(P')$, our choice of $s$ and $s'$ implies that $\mathtt{r}(\mathtt{QCprev}(P'))<s$. We reach a contradiction because $p_i$ sets its lock $\mathtt{Q}^+$ so that $\mathtt{r}(\mathtt{Q}^+)=s$ while in round $s$, and so would not vote for $P'$ in round $s'$ (line 15 of the pseudocode). 

\vspace{0.2cm} 
\noindent \textbf{The accumulation of small error probabilities.} The analysis is the same as in Section \ref{Snowmansecurity}, except that we must now account for two new assumptions on which we have conditioned in the argument above. 
Previously, we assumed that if at least 75\% of the correct processors have local $\mathtt{pref}$ values extending $\sigma$ at the beginning of round $s$, then at least 5/6 of the correct processors will have $\mathtt{pref}$ values extending $\sigma$ by the end of round $s$. Now, we require the additional assumption that if 5/6 of the correct processors have local $\mathtt{pref}$ values extending $\sigma$ at the beginning of round $s$, then at least 11/12 of the correct processors will have $\mathtt{pref}$ values extending $\sigma$ by the end of round $s$. For a given round $s$, a calculation for the binomial distribution shows that this holds, except with probability at most $2\times 10^{-47}$. If there are five rounds per second then, over a period of 1000 years, this means that less than $1.6\times 10^{11}$ rounds are executed. Applying the union bound, we conclude that this adds less than $4\times 10^{-36}$ to the cumulative error probability.

We must also account for the new decision condition. As noted previously, a calculation for the binomial distribution shows that if $p_i$ is correct then the probability that at least 3/5 of $p_i$'s sample sequence in a given round are Byzantine is less than $10^{-14}$. The probability that this happens in two given consecutive rounds is therefore less than $10^{-28}$. If at most 10,000 processors execute at most 5 rounds per second for 1000 years, this therefore adds less than $2\times 10^{-13}$ to the cumulative error probability. 

Overall, the same error bound of $3\times 10^{-5}$ that was established in Section \ref{Snowflakeanalysis} can be seen to hold here. 

\subsection{The proof of liveness}

Throughout this section, we assume that the value $E^*$ (specified in line 22 of the pseudocode for even epochs) is never empty for correct $p_i$, i.e. there are always new blocks to finalize. 

\vspace{0.2cm}
\noindent \textbf{Defining $\mathtt{final}_t$}. At any timeslot $t$, let $\mathtt{final}_t$ be the shortest amongst all local values $\mathtt{final}$ for correct processors (by Section \ref{consis} this value is uniquely defined, except with small  error probability). 

\vspace{0.2cm} The proof of liveness breaks into two parts: 

\vspace{0.1cm} 
\noindent \emph{Claim 3}. Suppose that all correct processors are in even epoch $e$ at $t$. Then, except with small  error probability, either $\mathtt{final}_{t+6\Delta \gamma}$ properly extends $\mathtt{final}_t$ or else all correct processors enter epoch $e+1$ by time $t+6 \Delta \gamma$. 

\vspace{0.1cm} 
\noindent \emph{Claim 4}. If all correct processors enter the odd epoch $e+1$, this epoch finalizes a new value. 

\vspace{0.1cm} In Claim 4, the number of rounds during epoch $e+1$ required to finalize a new value is  bounded by the maximum number of consecutive faulty leaders. Since we make the simple choice of using deterministic leader selection during odd epochs, this means that the number of required rounds is $O(f)$, but one could ensure the number of required rounds is $O(1)$ and maintain a small  chance of liveness failure by using random leader selection. 

\vspace{0.2cm}
\noindent \textbf{Establishing Claim 3}. Given the conditions in the statement of the claim, let $E^{\diamond}$ be the set of correct processors that have local $\mathtt{final}$ values properly extending $\mathtt{final}_t$ at time $t_1:=t+3\Delta \gamma$. If $|E^{\diamond}|\leq 3n/5$, then at least $n/5$ correct processors send epoch $e+1$ messages $(\text{stuck},e,\mathtt{final}_t)$ by time $t_1$, and all correct processors enter epoch $e+1$ by time $t_1 +\Delta$. So, suppose $|E^{\diamond}|> 3n/5$ and that it is not the case all correct processors enter epoch $e+1$ by time $t+6 \Delta \gamma$.  Let $x\in \{ 0, 1 \}$ be such that some correct processor finalizes $\mathtt{final}_t \ast x$, and (by consistency) condition on there existing a unique such $x$. Consider the instructions as locally defined for correct $p_i$ when executing any round $s$ of epoch $e$ that starts subsequent to $t_1$. A calculation for the binomial shows that the probability $|\{ j\in [1,k]: \mathtt{final}(j,s)\supseteq \mathtt{final}_t \ast x \}|\geq 48$ is at least 0.548. 
The probability that this holds in both of any two such consecutive rounds $s$ and $s+1$ is therefore at least 0.3. Since we suppose $\gamma \geq 300$, the probability that $p_i$ fails to finalize a value extending $\mathtt{final}_t$ by time  $t+6 \Delta \gamma$ is therefore at most $0.71^{150}<10^{-22}$. 

\vspace{0.2cm}
\noindent \textbf{Establishing Claim 4}. Towards a contradiction, suppose that all correct processors enter odd epoch $e+1$, but that the epoch never finalizes a new value. Let $t=3\Delta s$ be such that $\mathtt{lead}(s)$ is correct and all correct processors are in epoch $e+1$ at $t$, with their local value $\mathtt{ready}(e+1)$ equal to 1. Let $s'$ be the greatest such that any correct processor has a local value $\mathtt{Q}^+$ at $t$ with $\mathtt{r}(\mathtt{Q}^+)=s'$. Note that either $s'=0$, or else any correct processor that has set is local value  $\mathtt{Q}^+$ so that $\mathtt{r}(\mathtt{Q}^+)=s'$, did so at a timeslot $\leq t-\Delta$. According to our conventions regarding the gossiping of QCs, this means that $\mathtt{lead}(s)$ will receive a QC, $Q$ say, with $\mathtt{r}(Q)\geq s'$ by $t$. The correct processor $\mathtt{lead}(s)$ will then send out a proposal $P$ during round $s$ that will be regarded as an $\mathtt{M}$-valid proposal for round $s$ by all correct processors. If $\mathtt{Q}^+$ is as locally defined for any  correct processor at $t+\Delta$,  $P$  will also satisfy the condition that $\mathtt{r}(\mathtt{QCprev}(P))\geq \mathtt{r}(\mathtt{Q}^+)$. All correct processors will therefore send stage 1 and 2 votes for $P$, and $P$ will be confirmed for all correct processors. 

\vspace{0.2cm} 
\noindent \textbf{A comment on the finalization of blocks produced by correct leaders and the length of odd epochs}. For the sake of simplicity, we have structured the instructions for odd epochs so as to ensure the finalization of one more block, rather than so as to ensure the finalization of at least one more block produced by a correct leader. Of course, one could achieve the latter result simply by running odd epochs until at least $f+1$ distinct leaders have produced finalized blocks.

\section{Related work}

The Snow family of consensus protocols was introduced in \cite{rocket2019scalable}. Subsequent to this, Amores-Sesar, Cachin and Tedeschi \cite{amores2022spring} gave a complete description of the Avalanche protocol\footnote{The Avalanche protocol is a DAG-based variant of Snowman that does not aim to produce a total ordering on transactions, and was only described at a high level in \cite{rocket2019scalable}. It is not used in the present instantiation of the Avalanche blockchain.} and formally established security properties for that protocol, given an $O(\sqrt{n})$ adversary and assuming that the Snowball protocol (a variant of Snowflake$^+$) solves probabilistic Byzantine Agreement for such adversaries. The authors also described (and provided a solution for) a liveness attack. As noted in \cite{amores2022spring}, the original implementation of the Avalanche protocol used by the Avalanche blockchain (before replacing Avalanche with a version of Snowman that totally orders transactions) had already introduced modifications avoiding the possibility of such attacks. 

In \cite{amores2024analysis}, Amores-Sear, Cachin and Schneider consider the Slush protocol and show that coming close to a consensus already requires a minimum of $\Omega \left( \frac{\text{log }n}{\text{log }k} \right)$ rounds,
even in the absence of adversarial influence. They show that Slush reaches a stable consensus in $O(\text{log }n)$ rounds, and that this holds even when the adversary can influence up to $O(\sqrt{n})$ processors. They also show that the $\Omega \left( \frac{\text{log }n}{\text{log }k} \right)$ lower bound holds for Snowflake and Snowball. 

\vspace{0.2cm} 
There is a vast literature that considers a closely related family of models, from the \emph{Ising model} \cite{brush1967history} as studied in statistical mechanics, to \emph{voter models} \cite{holley1975ergodic} as studied in applied probability and other fields, to the \emph{Schelling model of segregation} \cite{schelling1969models} as studied by economists (and more recently by computer scientists \cite{brandt2012analysis,barmpalias2014digital} and physicists \cite{omidvar2017self,omidvar2021improved,ortega2021schelling}). Within this family of models there are many variants, but a standard approach is to consider a process that proceeds in rounds. In each round, each participant samples a small number of other participants to learn their present state, and then potentially updates their own state according to given rule. A fundamental difference with our analysis here is that, with two exceptions (mentioned below), such models do not incorporate the possibility of Byzantine action. Examples of such research aimed specifically at the task of reaching consensus include \cite{becchetti2016stabilizing,doerr2011stabilizing,elsasser2017brief,ghaffari2018nearly,cooper2014power,cruciani2021phase} (see \cite{becchetti2020consensus} for an overview). Amongst these papers, we are only aware of \cite{becchetti2016stabilizing} and \cite{doerr2011stabilizing} considering Byzantine action, and those two papers deal only with an $O(\sqrt{n})$ adversary. 

\vspace{0.2cm} FPC-BI \cite{popov2021fpc,popov2022voting} is a protocol which is closely related to the Snow family of consensus protocols, but which takes a different approach to the liveness issue (for adversaries which are larger than $O(\sqrt{n})$) than that described here. The basic idea behind their approach is to use a common random coin to dynamically and unpredictably set threshold parameters (akin to $\alpha_1$ and $\alpha_2$ here) for each round, making it much more difficult for an adversary to keep the honest population split on their preferred values. Since the use of a common random coin involves practical trade-offs, their approach and ours may be seen as complementary.

\section{Final comments} \label{fincom}

In this paper, we have considered the case that the adversary controls at most $f<n/5$ processors. We described the protocol Snowflake$^+$ and showed that it satisfies validity and agreement, except with small error probability. We showed how Snowflake$^+$ can be adapted to give an SMR protocol, Snowman, which satisfies consistency, except with small error probability. We then augmented Snowman with a liveness module, to form the protocol Frosty, which we proved satisfies liveness and consistencty except with small error probability. 
We note that Avalanche presently implements Snowflake, rather than Snowflake$^+$, and uses different parameters than those used in the proofs here. Snowflake$^+$ was implemented a few months prior to the writing of this paper, but is not yet activated. Error-driven Snowflake$^+$ is planned for implementation in the coming months. The community may consider adopting the parameters proposed in this paper because they provide a good tradeoff between consistency and latency.

\vspace{0.2cm}
In future work, we aim to expand the analysis here as follows: 
\begin{enumerate} 
\item[(i)] The bounds $f<n/5$ and $n\geq 500$ were used only so as to be able to give as simple a proof as possible in Section \ref{Snowflakeanalysis}. In subsequent papers, we intend to carry out a more fine-grained analysis for smaller $n$ and larger $f$.   
\item[(ii)] The analysis here was simplified by the assumption that processors execute instructions in synchronous rounds. In a follow-up work, we will show how the methods described here can be adapted to give formal proofs of consistency and liveness for a \emph{responsive} form of the protocol, allowing each processor to proceed individually through rounds as fast as network delays allow. 
\item[(iii)] While the liveness module described here achieves (probabilistic) liveness when $f<n/5$, we aim to explore ways in which \emph{slashing} can be implemented for liveness attacks. For $f<n/3$ this may be possible,  if one can show that liveness attacks require the adversary either to give provably false information to others, or else execute sampling that is provably biased.
\end{enumerate} 

\section{Acknowledgements} 
The authors would like to thank Christian Cachin, Philipp Schneider, and Ignacio Amores Sesar for a number of very useful conversations.

\bibliographystyle{plainurl}

\begin{thebibliography}{10}

\bibitem{abraham2019communication}
Ittai Abraham, TH~Hubert Chan, Danny Dolev, Kartik Nayak, Rafael Pass, Ling
  Ren, and Elaine Shi.
\newblock Communication complexity of byzantine agreement, revisited.
\newblock In {\em Proceedings of the 2019 ACM Symposium on Principles of
  Distributed Computing}, pages 317--326, 2019.

\bibitem{amores2024analysis}
Ignacio Amores-Sesar, Christian Cachin, and Philipp Schneider.
\newblock An analysis of avalanche consensus.
\newblock {\em arXiv preprint arXiv:2401.02811}, 2024.

\bibitem{amores2022spring}
Ignacio Amores-Sesar, Christian Cachin, and Enrico Tedeschi.
\newblock When is spring coming? a security analysis of avalanche consensus.
\newblock {\em arXiv preprint arXiv:2210.03423}, 2022.

\bibitem{barmpalias2014digital}
George Barmpalias, Richard Elwes, and Andy Lewis-Pye.
\newblock Digital morphogenesis via schelling segregation.
\newblock In {\em 2014 IEEE 55th Annual Symposium on Foundations of Computer
  Science}, pages 156--165. IEEE, 2014.

\bibitem{becchetti2020consensus}
Luca Becchetti, Andrea Clementi, and Emanuele Natale.
\newblock Consensus dynamics: An overview.
\newblock {\em ACM SIGACT News}, 51(1):58--104, 2020.

\bibitem{becchetti2016stabilizing}
Luca Becchetti, Andrea Clementi, Emanuele Natale, Francesco Pasquale, and Luca
  Trevisan.
\newblock Stabilizing consensus with many opinions.
\newblock In {\em Proceedings of the twenty-seventh annual ACM-SIAM symposium
  on Discrete algorithms}, pages 620--635. SIAM, 2016.

\bibitem{boneh2001short}
Dan Boneh, Ben Lynn, and Hovav Shacham.
\newblock Short signatures from the weil pairing.
\newblock In {\em International conference on the theory and application of
  cryptology and information security}, pages 514--532. Springer, 2001.

\bibitem{brandt2012analysis}
Christina Brandt, Nicole Immorlica, Gautam Kamath, and Robert Kleinberg.
\newblock An analysis of one-dimensional schelling segregation.
\newblock In {\em Proceedings of the forty-fourth annual ACM symposium on
  Theory of computing}, pages 789--804, 2012.

\bibitem{brush1967history}
Stephen~G Brush.
\newblock History of the lenz-ising model.
\newblock {\em Reviews of modern physics}, 39(4):883, 1967.

\bibitem{chen2016algorand}
Jing Chen and Silvio Micali.
\newblock Algorand.
\newblock {\em arXiv preprint arXiv:1607.01341}, 2016.

\bibitem{cooper2014power}
Colin Cooper, Robert Els{\"a}sser, and Tomasz Radzik.
\newblock The power of two choices in distributed voting.
\newblock In {\em International Colloquium on Automata, Languages, and
  Programming}, pages 435--446. Springer, 2014.

\bibitem{cruciani2021phase}
Emilio Cruciani, Hlafo~Alfie Mimun, Matteo Quattropani, and Sara Rizzo.
\newblock Phase transitions of the k-majority dynamics in a biased
  communication model.
\newblock In {\em Proceedings of the 22nd International Conference on
  Distributed Computing and Networking}, pages 146--155, 2021.

\bibitem{doerr2011stabilizing}
Benjamin Doerr, Leslie~Ann Goldberg, Lorenz Minder, Thomas Sauerwald, and
  Christian Scheideler.
\newblock Stabilizing consensus with the power of two choices.
\newblock In {\em Proceedings of the twenty-third annual ACM symposium on
  Parallelism in algorithms and architectures}, pages 149--158, 2011.

\bibitem{dolev1985bounds}
Danny Dolev and R{\"u}diger Reischuk.
\newblock Bounds on information exchange for byzantine agreement.
\newblock {\em Journal of the ACM (JACM)}, 32(1):191--204, 1985.

\bibitem{elsasser2017brief}
Robert Els{\"a}sser, Tom Friedetzky, Dominik Kaaser, Frederik Mallmann-Trenn,
  and Horst Trinker.
\newblock Brief announcement: rapid asynchronous plurality consensus.
\newblock In {\em Proceedings of the ACM symposium on principles of distributed
  computing}, pages 363--365, 2017.

\bibitem{garay2018bitcoin}
Juan~A Garay, Aggelos Kiayias, and Nikos Leonardos.
\newblock The bitcoin backbone protocol: Analysis and applications.
\newblock 2018.

\bibitem{ditto}
Rati Gelashvili,  Lefteris Kokoris-Kogias, Alberto Sonnino, Alexander Spiegelman, and Zhuolun Xiang. 
\newblock Jolteon and ditto: Network-adaptive efficient consensus with asynchronous fallback.
\newblock In {\em International conference on financial cryptography and data security}, pp. 296-315. Cham: Springer International Publishing, 2022.

\bibitem{ghaffari2018nearly}
Mohsen Ghaffari and Johannes Lengler.
\newblock Nearly-tight analysis for 2-choice and 3-majority consensus dynamics.
\newblock In {\em Proceedings of the 2018 ACM Symposium on Principles of
  Distributed Computing}, pages 305--313, 2018.

\bibitem{holley1975ergodic}
Richard~A Holley and Thomas~M Liggett.
\newblock Ergodic theorems for weakly interacting infinite systems and the
  voter model.
\newblock {\em The annals of probability}, pages 643--663, 1975.

\bibitem{king2011breaking}
Valerie King and Jared Saia.
\newblock Breaking the o (n 2) bit barrier: scalable byzantine agreement with
  an adaptive adversary.
\newblock {\em Journal of the ACM (JACM)}, 58(4):1--24, 2011.

\bibitem{lamport1982byzantine}
Leslie Lamport, Robert Shostak, and Marshall Pease.
\newblock The byzantine generals problem.
\newblock {\em ACM Transactions on Programming Languages and Systems (TOPLAS)},
  4(3):382--401, 1982.

\bibitem{lewis2023permissionless}
Andrew Lewis-Pye and Tim Roughgarden.
\newblock Permissionless consensus.
\newblock {\em arXiv preprint arXiv:2304.14701}, 2023.

\bibitem{plusplus}
Patrick O'Grady.
\newblock Apricot Phase Four: Snowman++ and Reduced C-Chain Transaction Fees.
\newblock \url{https://medium.com/avalancheavax/apricot-phase-four-snowman-and-reduced-c-chain-transaction-fees-1e1f67b42ecf}


\bibitem{omidvar2017self}
Hamed Omidvar and Massimo Franceschetti.
\newblock Self-organized segregation on the grid.
\newblock In {\em Proceedings of the ACM Symposium on Principles of Distributed
  Computing}, pages 401--410, 2017.

\bibitem{omidvar2021improved}
Hamed Omidvar and Massimo Franceschetti.
\newblock Improved intolerance intervals and size bounds for a schelling-type
  spin system.
\newblock {\em Journal of Statistical Mechanics: Theory and Experiment},
  2021(7):073302, 2021.

\bibitem{ortega2021schelling}
Diego Ortega, Javier Rodr{\'\i}guez-Laguna, and Elka Korutcheva.
\newblock A schelling model with a variable threshold in a closed city
  segregation model. analysis of the universality classes.
\newblock {\em Physica A: Statistical Mechanics and its Applications},
  574:126010, 2021.

\bibitem{pass2016hybrid}
Rafael Pass and Elaine Shi.
\newblock Hybrid consensus: Efficient consensus in the permissionless model.
\newblock {\em Cryptology ePrint Archive}, 2016.

\bibitem{popov2021fpc}
Serguei Popov and William~J Buchanan.
\newblock Fpc-bi: Fast probabilistic consensus within byzantine
  infrastructures.
\newblock {\em Journal of Parallel and Distributed Computing}, 147:77--86,
  2021.

\bibitem{popov2022voting}
Serguei Popov and Sebastian M{\"u}ller.
\newblock Voting-based probabilistic consensuses and their applications in
  distributed ledgers.
\newblock {\em Annals of Telecommunications}, pages 1--23, 2022.

\bibitem{rocket2019scalable}
Team Rocket, Maofan Yin, Kevin Sekniqi, Robbert van Renesse, and Emin~G{\"u}n
  Sirer.
\newblock Scalable and probabilistic leaderless bft consensus through
  metastability.
\newblock {\em arXiv preprint arXiv:1906.08936}, 2019.

\bibitem{schelling1969models}
Thomas~C Schelling.
\newblock Models of segregation.
\newblock {\em The American economic review}, 59(2):488--493, 1969.

\bibitem{schneider1990implementing}
Fred~B Schneider.
\newblock Implementing fault-tolerant services using the state machine
  approach: A tutorial.
\newblock {\em ACM Computing Surveys (CSUR)}, 22(4):299--319, 1990.

\bibitem{shoup2000practical}
Victor Shoup.
\newblock Practical threshold signatures.
\newblock In {\em Advances in Cryptology—EUROCRYPT 2000: International
  Conference on the Theory and Application of Cryptographic Techniques Bruges,
  Belgium, May 14--18, 2000 Proceedings 19}, pages 207--220. Springer, 2000.

\end{thebibliography}

\end{document}